\documentclass[lettersize,journal]{IEEEtran}
\IEEEoverridecommandlockouts
\usepackage{cite}
\usepackage{amsmath,amssymb,amsfonts}
\usepackage{algorithmic}
\usepackage{graphicx}
\usepackage{textcomp}
\usepackage{xcolor}
\usepackage{hyperref}
\usepackage{booktabs}
\usepackage{mathrsfs}
\usepackage{diagbox}
\usepackage{tabu}
\usepackage{stfloats}
\usepackage{caption}
\usepackage{multirow}
\usepackage{url}
\usepackage{ulem}
\usepackage{utfsym}
\usepackage{arydshln}
\usepackage[T1]{fontenc}
\usepackage{graphicx}
\usepackage{float}
\usepackage{subfig}

\makeatletter
\newcommand{\linebreakand}{%
  \end{@IEEEauthorhalign}
  \hfill\mbox{}\par
  \mbox{}\hfill\begin{@IEEEauthorhalign}
}
\makeatother

\def\BibTeX{{\rm B\kern-.05em{\sc i\kern-.025em b}\kern-.08em
    T\kern-.1667em\lower.7ex\hbox{E}\kern-.125emX}}
\begin{document}

\title{LogLLM: Log-based Anomaly Detection Using Large Language Models
}





\author{Wei Guan,
        Jian Cao,
        Shiyou Qian,
        Jianqi Gao,
        Chun Ouyang
\IEEEcompsocitemizethanks{
\IEEEcompsocthanksitem  W. Guan is with the School of Computer Software, Tianjin University, Tianjin, China. E-mail: guan\_wei@tju.edu.cn.
\IEEEcompsocthanksitem J. Cao, S. Qian are with the School of Computer Science, Shanghai Jiao Tong University, Shanghai, China. E-mail:  \{cao-jian, qshiyou\}@sjtu.edu.cn.
\IEEEcompsocthanksitem 
J. Gao is with the School of Computer Engineering and Science, Shanghai University, China. 
E-mail: jianqi\_gao@shu.edu.cn.
\IEEEcompsocthanksitem C. Ouyang is with the School of Information Systems, Queensland University of Technology, Brisbane, Australia. Email: c.ouyang@qut.edu.au
}
\thanks{Corresponding author: J. Cao.}
}

\IEEEtitleabstractindextext{%
\begin{abstract}
Software systems often record important runtime information in logs to help with troubleshooting. Log-based anomaly detection has become a key research area that aims to identify system issues through log data, ultimately enhancing the reliability of software systems. 
Existing methods often fall short in capturing the semantic information, typically expressed in natural language, or the sequential dependencies inherent in log sequences.
In this paper, we propose LogLLM, a framework that enables collaboration between heterogeneous LLMs for log-based anomaly detection.
LogLLM exploits the complementary capabilities of different LLM architectures: a Transformer encoder-based LLM is employed to extract fine-grained semantic vectors from individual log messages, while a Transformer decoder-based LLM is utilized to model sequential dependencies and generate anomaly detection decisions.
To enable effective collaboration between these heterogeneous LLMs, we introduce a learnable projector to align their vector representation spaces.
Furthermore, we design a progressive three-stage training strategy to optimize the collaboration between heterogeneous LLMs by gradually aligning their representations and adapting them to log anomaly detection.
Unlike conventional methods that require log parsers to extract templates, LogLLM preprocesses log messages with regular expressions, streamlining the entire process.
Experimental results on four public real-world datasets demonstrate that LogLLM outperforms state-of-the-art methods, achieving an average F$_1$-score improvement of 6.6\% over the strongest existing approach. 
Further analyses provide insights into the effectiveness of the key components of the model architecture and progressive three-stage training strategy.
\end{abstract}

}

\maketitle

\IEEEdisplaynontitleabstractindextext

\IEEEpeerreviewmaketitle

\section{Introduction}
\IEEEPARstart{E}{n}suring high availability and reliability is crucial for large-scale software-intensive systems \cite{albert2025system}. As these systems become more complex and expansive, the occurrence of anomalies becomes unavoidable \cite{le2021log, 10738505}. Even a minor issue can lead to system instability, compromised reliability, data integrity problems, and substantial losses in revenue and customer trust.
Therefore, anomaly detection is vital for maintaining the health and stability of complex software-intensive systems \cite{zhang2024end,studiawan2020anomaly,skopik2022behavior}, which typically produce system logs that record system states and critical runtime events \cite{le2022log}. 
Engineers can utilize this log data to evaluate system health, identify anomalies, and trace the root causes of issues.
However, due to the potentially vast volume of logs, manually analyzing them for anomalies can be both labor-intensive and prone to mistakes \cite{qi2023loggpt}. 
Consequently, log-based anomaly detection has emerged as a key area in automated log analysis, focusing on the automatic identification of system anomalies through log data.

Numerous deep learning-based methods \cite{du2017deeplog,meng2019loganomaly,zhang2021logattn, catillo2022autolog, xie2023log, zhang2023anomaly,duan2021generative,he2023graph,zhang2023layerlog,hashemi2024onelog, lu2018detecting,zhang2019robust,xie2022loggd, yilmaz2025novel,zhao2022trine,yang2021semi} for log-based anomaly detection have been proposed. These methods typically employ sequential deep learning models such as LSTM \cite{hochreiter1997long} and Transformers \cite{vaswani2017attention}. These methods can be further divided into reconstruction-based methods   and binary classification-based methods. Reconstruction-based methods \cite{du2017deeplog,meng2019loganomaly,zhang2021logattn, catillo2022autolog, xie2023log, zhang2023anomaly,duan2021generative,he2023graph} involve designing and training a deep neural network to reconstruct input log sequences, with anomalies detected based on reconstruction errors. The underlying principle is that anomalous samples cannot be accurately reconstructed. Binary classification-based methods \cite{zhang2023layerlog,hashemi2024onelog, lu2018detecting,zhang2019robust,xie2022loggd,  yilmaz2025novel, zhao2022trine,yang2021semi}, on the other hand, involve designing a binary classifier to classify samples as either normal or anomalous. These methods often require labeled anomalies for training purposes.
It is recognized that system logs are documented in natural language and contain a significant amount of semantic information. 
Nevertheless, traditional deep learning-based methods commonly represent log messages as discrete IDs, neglecting the semantic information encoded in natural language. Consequently, they are less robust to unseen or evolving log templates.

In recent years, significant advancements have been achieved in Large Language Models (LLMs). Existing LLMs can be categorized into \textit{Transformer encoder-based LLMs}, such as BERT \cite{devlin2018bert}, RoBERTa \cite{liu2019roberta}, and SpanBERT \cite{joshi2020spanbert}, and \textit{Transformer decoder-based LLMs}, including Llama 3 \cite{dubey2024llama}, Qwen3 \cite{yang2025qwen3}, and GLM-4 \cite{glm2024chatglm}.
These models are characterized by their vast parameter sizes and are pretrained on substantially larger datasets, ranging from several gigabytes to terabytes in size. 
This extensive pretraining equips them with remarkable language comprehension abilities, enabling superior performance in tasks such as summarization, paraphrasing, and instruction following, even in zero-shot scenarios \cite{guan2024dabl}.
Existing methods that utilize LLMs for log-based anomaly detection can be categorized into training-free methods and fine-tuning-based methods.
Training-free methods \cite{liu2024logprompt,qi2023loggpt,egersdoerfer2023early,pan2024raglog,zhang2025xraglog,duan2025eagerlog, zhang2025scalalog,sui2025bridging,no2024training} perform anomaly detection using LLMs without task-specific model adaptation. 
However, they often struggle to capture dataset-specific characteristics, which may lead to suboptimal detection performance.
Fine-tuning-based methods \cite{guo2021logbert,lee2023lanobert,lin2024fastlogad,almodovar2024logfit,le2021log,  chen2022bert, adeba2024sarlog,fu2023mlog, chu2025anomaly,hadadi2024anomaly,yang2025logllama} integrate LLMs into deep neural networks and tailor them to user-specific datasets. Nevertheless, these methods encounter challenges such as limited semantic understanding, suboptimal LLM utilization (relying solely on LLMs for semantic information extraction), and insufficient consideration of input data format, which can lead to memory overflow, as discussed in Section \ref{sec:llmad}.

To tackle the aforementioned challenges, we propose LogLLM, a fine-tuning-based method that enables collaboration between heterogeneous LLMs for log-based anomaly detection. 
Unlike existing approaches that typically rely on a single type of LLM, LogLLM exploits the complementary capabilities of different LLM architectures. 
Specifically, a Transformer encoder-based LLM is employed to extract fine-grained semantic representations from individual log messages, while a Transformer decoder-based LLM is utilized to model sequential dependencies and generate anomaly detection decisions. 
Since these heterogeneous LLMs have different representation spaces, directly integrating them is non-trivial. 
To address this issue, we introduce a learnable projector that aligns their representation spaces, enabling effective interaction between the Transformer encoder-based and decoder-based LLMs. 
Furthermore, we design a progressive three-stage training strategy to optimize the collaboration between heterogeneous LLMs. Instead of directly fine-tuning all model components simultaneously, the proposed strategy first optimizes the Transformer decoder-based LLM for anomaly response generation, then optimizes the Transformer encoder-based LLM and projector to generate semantic representations aligned with the representation space of the Transformer decoder-based LLM, and finally jointly fine-tunes the integrated model.
Unlike traditional methods that rely on log parsers to extract templates by identifying constant patterns, LogLLM preprocesses log messages using Regular Expressions (REs) to identify and replace variable parts, alleviating the limitations of log parsers in extracting unseen log templates.


\begin{figure*}[tb]
	\centering
    \includegraphics[scale=0.6]{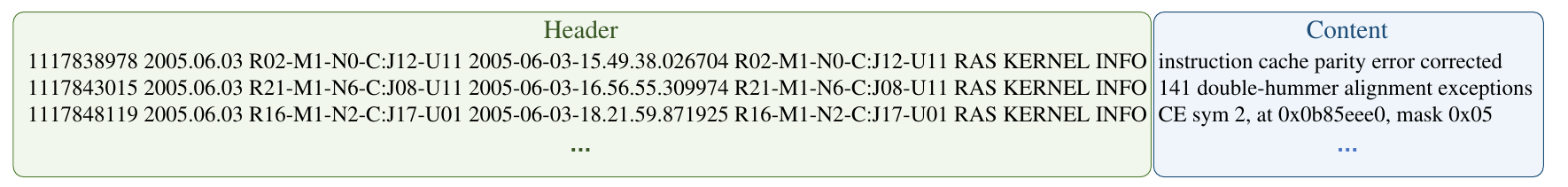}
 	\caption{An example of a raw system log from the BGL dataset, consisting of header and content fields.}
	\label{fig:log}
\end{figure*}

Transformer decoder-based LLMs frequently face out-of-memory challenges due to their extensive parameter sizes \cite{burtsev2024working,hadadi2024anomaly}.  
As discussed in Section \ref{sec:ch}, directly feeding an entire log sequence (by concatenating log messages into a single long string) into a Transformer decoder-based LLM not only causes out-of-memory issues but also blurs the boundaries between individual messages. This ambiguity makes it more difficult for the model to capture meaningful sequential dependencies and identify critical patterns for anomaly detection.
LogLLM effectively addresses these limitations by employing a Transformer encoder-based LLM to encode each log message, which significantly reduces the number of embedding vectors fed into the decoder-based LLM and clearly delineates the boundaries between individual log messages (one message per embedding vector), thereby enhancing anomaly detection performance.
We conduct experiments across four public datasets, and the results demonstrate that LogLLM outperforms state-of-the-art methods,  achieving an average F$_1$-score improvement of 6.6\% over the strongest existing approach. 
Further analyses validate the effectiveness of the key components of the model architecture and the progressive three-stage training strategy.
The main contributions of our work are as follows:
\begin{itemize}
    \item 
    We introduce LogLLM, a heterogeneous LLM collaboration framework that integrates Transformer encoder-based and decoder-based LLMs to capture log message-level semantics and sequence-level dependencies for log-based anomaly detection.
    \item 
    We propose a learnable representation alignment mechanism and a progressive three-stage training strategy to facilitate effective collaboration between heterogeneous LLMs.
    \item
    We conduct extensive experiments on four publicly available real-world datasets, demonstrating that LogLLM achieves exceptional performance.
\end{itemize}

\section{Background and Preliminaries}
\label{sec:Preliminaries}
To establish the groundwork for subsequent sections, we introduce the \textbf{system log}, which records the system’s events  and internal states during runtime.  A system log contains a list of log messages in chronological order.

Fig. \ref{fig:log} presents a snippet of a raw system log generated by the BGL (the BlueGene/L supercomputer system), with each log message ordered according to the recorded time.
These raw log messages are semi-structured texts consisting of a \textbf{header} and \textbf{content}. The header, determined by the logging framework, includes information such as timestamp, verbosity level (e.g., WARN/INFO), and component \cite{zhu2019tools}. The log content comprises a constant part (keywords that reveal the log template) and a variable part (parameters that carry dynamic runtime information). Following the convention of log-based anomaly detection methods, this paper focuses solely on the content of each log message.

\begin{figure*}[t]
    	\centering
    	\vskip -0.1in
    \subfloat[Session window]{
	\centering
        \includegraphics[scale=0.55]{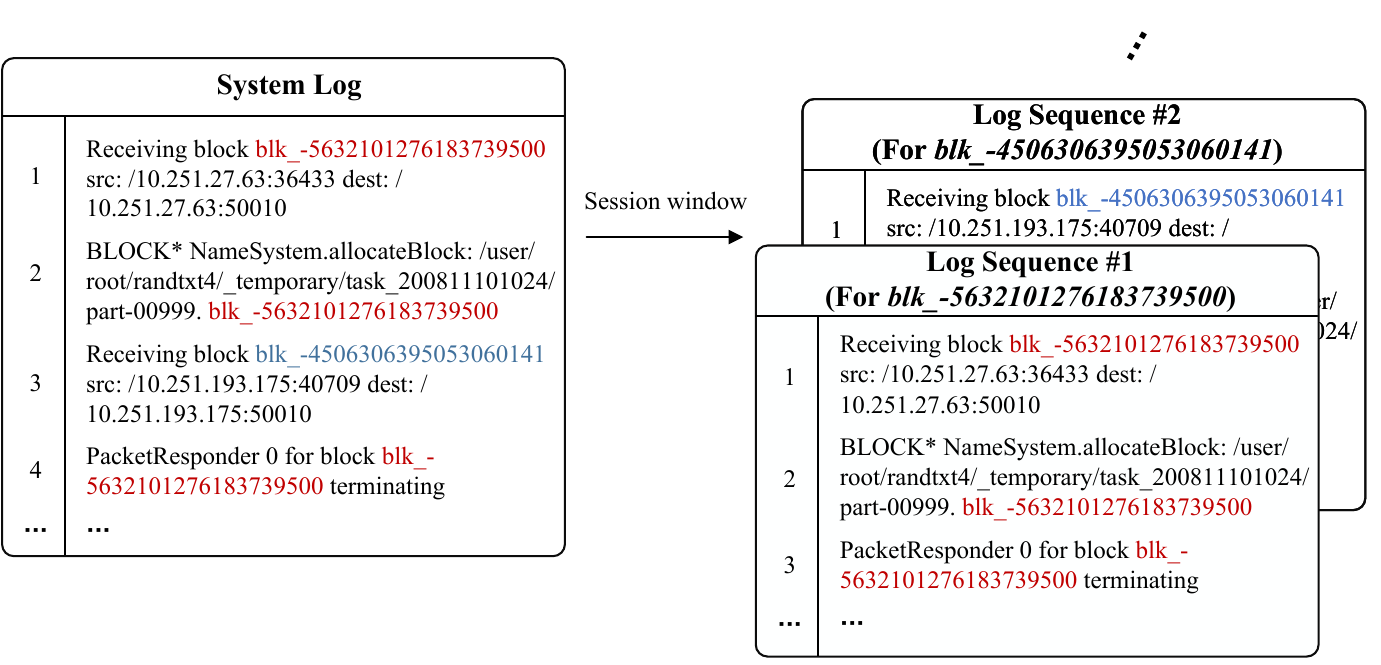}
	\label{fig:Session}
	}
	\hfil
\subfloat[Sliding window]{
	\centering
    \includegraphics[scale=0.55]{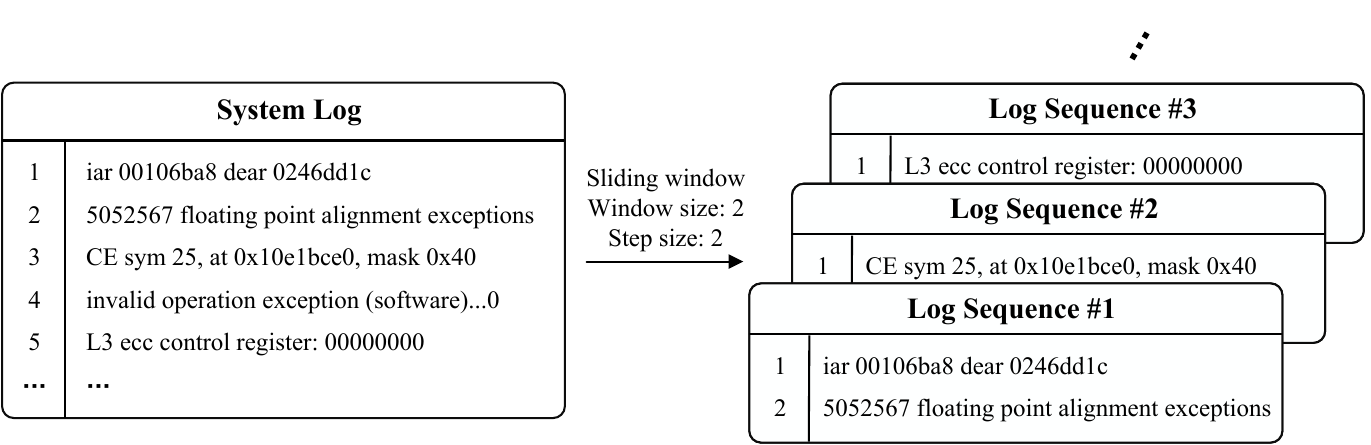}
	\label{fig:Fixed}
	} 
	\caption{Illustrative examples of log message partitioning.}
	\vskip -0.1in
\end{figure*}

The log messages can be grouped into \textbf{log sequences} (i.e., series of log messages that record specific execution flows) based on session or fixed/sliding windows \cite{he2016evaluation}. 
\textbf{Session window} partitioning groups log messages according to their session IDs,  thereby generating sequences that include the log messages within each session. For example, Fig. \ref{fig:Session}  illustrates the HDFS \cite{xu2009online} logs undergoing the session window grouping process, where the \textit{block\_id} serves as the session ID.
In contrast, \textbf{fixed/sliding window} partitioning groups log messages based on a fixed size (window size), which can be defined by either the time span or the number of log messages. This method creates sequences that capture snapshots of system log messages over time. 
For example, Fig. \ref{fig:Fixed}  illustrates the BGL  \cite{oliner2007supercomputers} logs undergoing the sliding window grouping process,  with a window size of 2 messages and a step size of 2 messages.

The objective of \textbf{log-based anomaly detection} in this paper is to train a model to identify anomalous log sequences, thereby facilitating the recognition of potential issues in the system’s operational behavior.
The model is trained using Supervised Fine-Tuning (SFT), with anomalies defined according to the labeled dataset. 
Specifically, whether a log sequence is anomalous depends on both the semantic information conveyed by individual log messages and the sequential dependencies among them. These two aspects are complementary rather than mutually exclusive. A \textbf{semantic anomaly} refers to a situation where the semantic information conveyed by log messages deviates from normal system behaviors, whereas a \textbf{sequential anomaly} refers to a situation where the contextual relationships or ordering patterns among log messages differ from normal execution patterns. Therefore, the anomaly of a log sequence is jointly determined by the semantic characteristics of individual messages and their contextual dependencies within the sequence.

\section{Methodology}
\label{sec:method}
\begin{figure*}[tb]
	\centering
    \includegraphics[scale=0.5]{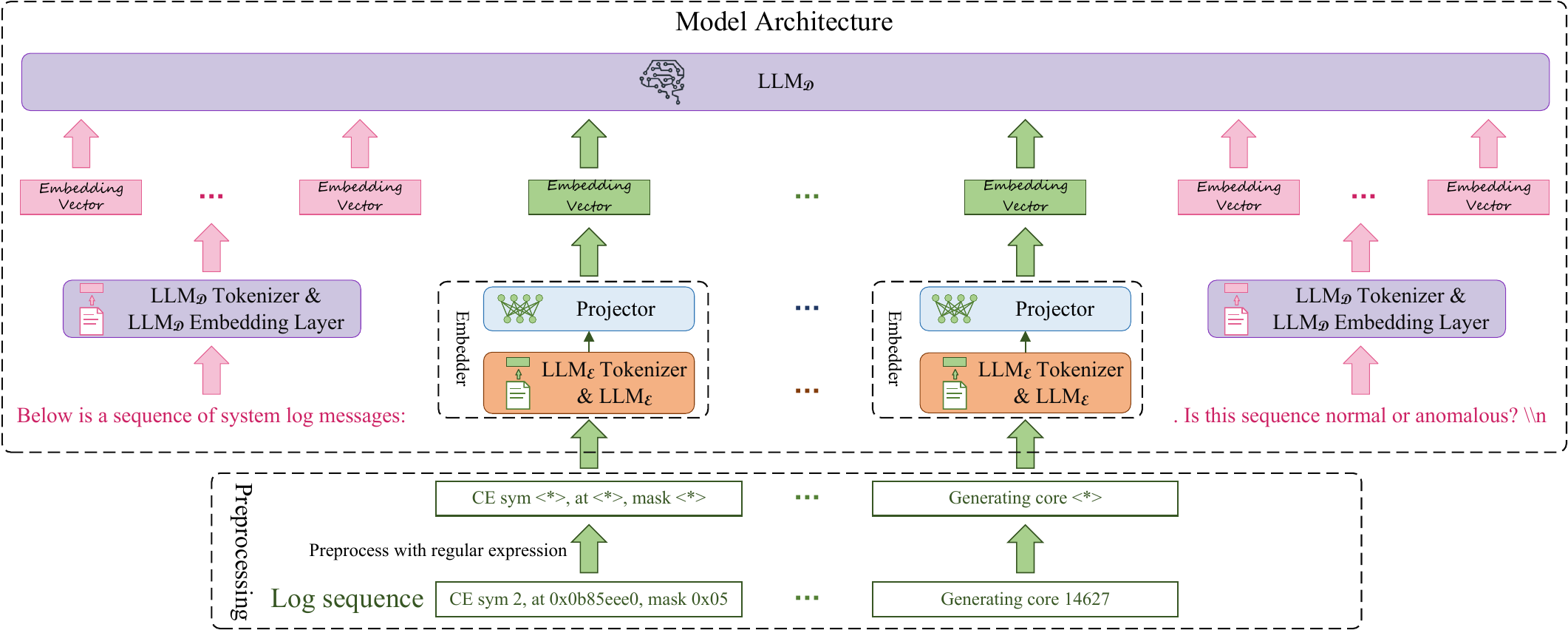}
    \caption{Overview of the proposed LogLLM framework. Log sequences are first preprocessed using regular expressions, then encoded into embedding vectors by the embedder, composed of LLM$_\mathcal{E}$ and the projector. The embedding vectors of the log sequence and textual prompts are fed into LLM$_\mathcal{D}$, which predicts whether a log sequence is anomalous. Notably, the model includes a single instance of LLM$_\mathcal{E}$ and the projector.}
	\label{fig:MA}
\end{figure*}

In this section, we present our innovative log-based anomaly detection framework, LogLLM.
As illustrated in Fig. \ref{fig:MA}, the log sequence undergoes preprocessing using regular expressions before being fed into a deep neural network that integrates a Transformer encoder-based LLM, a projector, and a Transformer decoder-based LLM for log-based anomaly detection. In the following sections, we will provide detailed insights into log sequence preprocessing, the architecture of the deep model, and the training strategy.

\subsection{Preprocessing}
\label{sec:Preprocessing}
Log messages typically consist of constant parts that describe logging events and variable parts that represent dynamic runtime information. 
Variable parts may introduce irrelevant variations that complicate deep model training, as demonstrated in Section \ref{sec:effectpre}. Therefore, accurately identifying and replacing them is essential for robust log-based anomaly detection.
Existing log-based anomaly detection methods commonly rely on log parsers, such as Drain \cite{he2017drain} and Spell \cite{du2016spell}, to extract log templates by identifying constant patterns in log messages. 
However, as noted by Le et al. \cite{le2021log}, existing log parsers do not always perform correctly on all log datasets and struggle to handle Out-Of-Vocabulary (OOV) words in new log messages. 
When logs are unstable, meaning new log templates emerge due to software evolution, these parsers gradually become less effective, making it increasingly difficult to support subsequent anomaly detection.

Fortunately, due to the structured generation process of system logs, many variable parameters corresponding to runtime-specific information can be identified through manually designed regular expressions \cite{le2023log}. 
Therefore, LogLLM directly identifies and replaces each variable parameter, such as account, directory path, and IP address, with a unified token `<*>'. 
Despite its simplicity, this preprocessing technique offers significant performance advantages.  Compared with existing log parsers, this strategy is more effective and does not require training, as demonstrated in Section \ref{sec:effectpre}.

\subsection{Model Architecture}
\label{sec:MA}
As shown in Fig. \ref{fig:MA}, our deep model consists of three main components: LLM$_\mathcal{E}$, implemented with a Transformer encoder-based LLM, a projector, and LLM$_\mathcal{D}$,  implemented with a Transformer decoder-based LLM.
LLM$_\mathcal{E}$ is utilized to extract vector representations of log messages, while  LLM$_\mathcal{D}$ is employed to predict whether the given log sequence is anomalous. The projector serves as a bridge, aligning the vector representation spaces of  LLM$_\mathcal{E}$ and LLM$_\mathcal{D}$.
It is important to note that our model incorporates only one instance of LLM$_\mathcal{E}$ and one projector.

\subsubsection{LLM$_\mathcal{E}$} Transformer encoder-based LLMs, such as BERT \cite{devlin2018bert}, RoBERTa \cite{liu2019roberta}, and SpanBERT \cite{joshi2020spanbert}, are designed to capture contextual dependencies within an input text by leveraging self-attention mechanisms that model relationships among tokens in a bidirectional manner. Compared to Transformer decoder-based LLMs, encoder-based LLMs are particularly effective at producing fixed-length vector representations that encapsulate the semantic meaning of the input text.
LLM$_\mathcal{E}$ serves as the semantic feature extractor in our framework.
Each log message, once preprocessed, is encoded into a semantic vector through the tokenizer and model of LLM$_\mathcal{E}$. For a preprocessed log sequence, the output of LLM$_\mathcal{E}$ is a sequence of semantic vectors $C=(c_1,c_2,\dots, c_N)\in \mathbb{R}^{N\times d_{\text{LLM}_\mathcal{E}}}$, where $N$ represents the length of the log sequence (i.e., the number of log messages) and $d_{\text{LLM}_\mathcal{E}}$ is the dimensionality  of each semantic vector (i.e., hidden size).  

\subsubsection{Projector}
The projector is a linear layer that maps the semantic vectors $C\in \mathbb{R}^{N\times d_{\text{LLM}_\mathcal{E}}}$ generated by LLM$_\mathcal{E}$ to the token embedding vectors accepted by LLM$_\mathcal{D}$, represented as $E=(e_1,e_2, \dots ,e_N) \in \mathbb{R}^{N\times d_{\text{LLM}_\mathcal{D}}}$, where $d_{\text{LLM}_\mathcal{D}}$  is the hidden size of LLM$_\mathcal{D}$.
The projector is designed to align the vector representation spaces of LLM$_\mathcal{E}$ and LLM$_\mathcal{D}$.

\subsubsection{LLM$_\mathcal{D}$}
Transformer decoder-based LLMs,  including Llama 3 \cite{dubey2024llama}, Qwen3 \cite{yang2025qwen3}, and GLM-4 \cite{glm2024chatglm}, are designed to generate coherent and contextually relevant text by predicting the next token in a sequence based on preceding tokens through a unidirectional self-attention mechanism. Compared to Transformer encoder-based LLMs, decoder-based LLMs excel at modeling sequential dependencies and generating fluent text.
LLM$_\mathcal{D}$ is employed to predict whether the given log sequence  is anomalous. To conduct supervised fine-tuning on LLM$_\mathcal{D}$, we generate corresponding textual queries based on embedded log sequences. Specifically, each query consists of three components: 
The first component introduces the log sequence, such as ``\textit{Below is a sequence of system log messages:}''. 
The second component comprises the embedding vectors $E$ output by the projector.
The third component queries whether the sequence is anomalous, asking, for instance, ``\textit{. Is this sequence normal or anomalous?}''. 
The first and third components are fed into the tokenizer and embedding layer of LLM$_\mathcal{D}$ sequentially, producing $E_{1}\in \mathbb{R}^{A\times d_{\text{LLM}_\mathcal{D}}}$ and $E_{3}\in \mathbb{R}^{Q\times d_{\text{LLM}_\mathcal{D}}}$, where $A$ and $Q$ are the number of tokens produced by tokenizing the first and third components, respectively.
Then, the embedding vectors of the three components are concatenated along the sequence dimension, represented as $[E_1|| E || E_3] \in  \mathbb{R}^{(A+N+Q)\times d_{\text{LLM}_\mathcal{D}}}$ and fed into  LLM$_\mathcal{D}$. 

\subsection{Training}
\label{sec:Training}
\subsubsection{Minority Class Oversampling}
\label{sec:mco}
LogLLM is a supervised anomaly detection method, which means it needs labeled normal and anomalous samples for training. However, supervised anomaly detection methods often face the challenge of data imbalance, which can lead to biased model training.  In an anomaly detection task, there are only two classes: normal and anomalous, and the number of instances in each class is uncertain. To cope with data imbalance, we oversample the class with fewer samples, ensuring that the proportion of the minority class is no less than $\beta$.
Formally, let the proportion of the minority class be $\alpha$ and  $\alpha <\beta$, and the total number of samples be $N_{sample}$.
To achieve a proportion of $\beta$ for the minority class, it will be oversampled to the following quantity:
\begin{equation}
\label{eq:oversample}
    \frac{\beta(1-\alpha)}{1-\beta}\times N_{sample} 
\end{equation}
This adjustment will make the proportion of the minority class equal to $\beta$.

\subsubsection{Training Objective}
Our objective is to train the deep model to determine whether a given log sequence is normal or anomalous. We fine-tune the model to generate a textual response accordingly: if the sequence is anomalous, it outputs ``\textit{The sequence is anomalous.}''; if normal, it outputs ``\textit{The sequence is normal.}''. We utilize cross-entropy loss \cite{bridle1990probabilistic} as our loss function.

\subsubsection{Training Procedure}
To train our deep model, we adopt a progressive three-stage training strategy designed to enhance the model's understanding of log sequences and improve its anomaly detection capability. Each stage builds upon the previous one to ensure both the components and the integrated model are optimized effectively.

\textbf{Stage 1. Fine-tuning LLM$_\mathcal{D}$ for response format adaptation: } 
In the first stage, we focus on fine-tuning the LLM$_\mathcal{D}$ for response format adaptation. 
Specifically, we train LLM$_\mathcal{D}$ to respond to the prompt ``\textit{Is this sequence normal or anomalous?}'' with ``\textit{The sequence is anomalous/normal.}''. 
To achieve this, we freeze the model's embedder, including LLM$_\mathcal{E}$ and the projector, while activating LLM$_\mathcal{D}$ for fine-tuning. 
By doing so, the model learns the desired response format efficiently without altering the pre-trained knowledge in LLM$_\mathcal{E}$. This stage requires only a small number of annotated samples.
Without this stage, the embedder may be misled, causing it to produce incorrect semantic representations of log messages and potentially resulting in model failure, as demonstrated in Section \ref{sec:exptp}.

\textbf{Stage 2. Training the embedder of log messages: } 
The second stage focuses on training the embedder to generate high-quality embeddings of log messages, which LLM$_\mathcal{D}$ can then use to accurately discern anomalies. 
During this stage, LLM$_\mathcal{D}$ is frozen, while the projector and LLM$_\mathcal{E}$ are activated for fine-tuning. This configuration allows the model to specialize in log-specific embeddings while preserving the general contextual knowledge already captured by LLM$_\mathcal{D}$. By optimizing the embedder in this way, the embeddings become better aligned with LLM$_\mathcal{D}$’s embedding space, facilitating more accurate anomaly classification in the subsequent full-model fine-tuning stage.

\textbf{Stage 3. Fine-tuning the entire model: } 
In the final stage, we perform end-to-end fine-tuning of the integrated model. This strategy allows the model to jointly optimize embedding generation and anomaly classification, ensuring cohesive performance across all components.


\section{Experiments}
\label{sec:exp}
In this section, we conduct extensive experiments on four real-life logs to investigate the following Research Questions (RQs):
\begin{itemize}
    \item \textbf{RQ1}: How effective is LogLLM in log-based anomaly
    detection?
    \item \textbf{RQ2}: How do different preprocessing techniques impact the performance of LogLLM?
    \item \textbf{RQ3}: How effective is the embedder for LLM$_\mathcal{D}$?
    \item \textbf{RQ4}: How do the design choices of LLM$_\mathcal{D}$ affect the performance and efficiency of LogLLM? 
    \item \textbf{RQ5}: How does each stage of the progressive three-stage training strategy affect the performance of LogLLM?
    \item \textbf{RQ6}: How does Minority Class Oversampling (MCO) affect the performance of LogLLM?
    
\end{itemize}

LogLLM is coded in Python, and the source code is available at  \href{https://github.com/guanwei49/LogLLM}{https://github.com/guanwei49/LogLLM}.

\subsection{Benchmark Methods}
To verify the superiority of the proposed method, we compare LogLLM with six state-of-the-art semi-supervised methods: DeepLog \cite{du2017deeplog}, LogAnomaly \cite{meng2019loganomaly}, PLELog \cite{yang2021semi}, FastLogAD \cite{lin2024fastlogad}, LogBERT \cite{guo2021logbert} and LogLLaMA \cite{yang2025logllama}.
We also compare it with three supervised methods: LogRobust \cite{zhang2019robust}, CNN \cite{lu2018detecting} and  NeuralLog \cite{le2021log}.
In addition, we include two training-free retrieval-based methods that do not require model training but rely on a retrieval corpus for anomaly detection: RAPID \cite{no2024training} and EagerLog \cite{duan2025eagerlog}.
For EagerLog, we adopt the \textit{Qwen3-8B} model\footnote{https://huggingface.co/Qwen/Qwen3-8B} as the LLM backbone instead of using an API-based model to reduce inference latency.
Additionally, we introduce a custom baseline, BGE-Trans, to verify the necessity of the proposed architecture. BGE-Trans employs \textit{bge-base-en-v1.5} model\footnote{https://huggingface.co/BAAI/bge-base-en-v1.5} to encode log messages into semantic vectors, which are then fed into a two-layer Transformer decoder with positional encoding; the final hidden state of the last token is passed to a linear layer for binary classification.

Among these baselines, FastLogAD, LogBERT, NeuralLog, EagerLog, LogLLaMA, RAPID and BGE-Trans leverage LLMs for log anomaly detection. Specifically, RAPID and EagerLog are training-free methods, whereas FastLogAD, LogBERT, NeuralLog, LogLLaMA and BGE-Trans are fine-tuning-based methods.


\subsection{Experimental Settings}
We conduct all experiments on a server equipped with an Intel Xeon Gold 6330 CPU (28 cores), 256GB of memory, and an NVIDIA A40 GPU with  48 GB of memory.

For the preprocessing step, we manually design a general and dataset-independent set of regular-expression rules to identify common runtime-specific information in log messages, including boolean values, numbers or textual numbers, date, IP/MAC addresses, file paths, hexadecimal identifiers, email/host identifiers, and alphanumeric tokens containing digits. The same set of RE rules is applied to all datasets without any dataset-specific modification.

Regarding model configurations, unless otherwise specified, we use the \textit{BERT-base} model\footnote{https://huggingface.co/google-bert/bert-base-uncased} as LLM$_\mathcal{E}$ and the \textit{Llama-3-8B} model\footnote{https://huggingface.co/meta-llama/Meta-Llama-3-8B} as LLM$_\mathcal{D}$. Specifically, BERT-base has a hidden size of 768, 12 hidden layers, and supports a maximum sequence length of 512 tokens. Llama-3-8B, on the other hand, has a hidden size of 4096, 32 hidden layers, and supports a maximum sequence length of 8192 tokens. 
To reduce the costs involved in fine-tuning LLMs with a substantial number of parameters, we utilize QLoRA \cite{dettmers2023qlora} to minimize memory usage. QLoRA accomplishes this by backpropagating gradients into a frozen 4-bit quantized model, while maintaining the performance levels achieved during the full 16-bit fine-tuning process. Specifically, we set $lora\_rank$ to $4$ and $lora\_alpha$ to $32$ for the BERT-base model and $lora\_rank$ to $8$ and $lora\_alpha$ to $16$ for the Llama-3-8B model.
The hyperparameter $\beta$, which is described in Section \ref{sec:mco}, is set to 30\%. We use the AdamW optimizer \cite{loshchilov2017decoupled} to train the model with a mini-batch size of 16. Unless otherwise specified, the training procedure is configured as follows: In the first stage, only 1,000 samples are involved with a learning rate of 5e-4. The second and third stages each consist of two epochs with a learning rate of 5e-5.

For a fair comparison, we use the publicly released source code provided by the original authors for all compared methods and configure the hyperparameters according to the values reported in their original papers.


\begin{table*}[t]\renewcommand{\arraystretch}{}
\setlength{\tabcolsep}{1.6mm}{
\caption{The statistics of datasets used in the experiments.}
\label{tab:stastic}
\begin{tabular}{ccccccccc}
\toprule
    & \multirow{2}{*}{\# Log messages} & \multirow{2}{*}{\# Log sequences} & \multicolumn{3}{c}{Training Data}               & \multicolumn{3}{c}{Testing Data}                \\ \cmidrule(lr){4-6} \cmidrule(lr){7-9} 
            &                                  &                                   & \# Log sequences & \# Anomalies & Anomaly ratio & \# Log sequences & \# Anomalies & Anomaly ratio \\ \midrule
HDFS        & 11,175,629                       & 575,061                           & 460,048          & 13,497        & 2.93\%         & 115,013           & 3,341         & 2.90\%         \\
BGL         & 4,747,963                        & 47,135                            & 37,708           & 4,009         & 10.63\%        & 9,427             & 817          & 8.67\%         \\
Liberty     & 5,000,000                        & 50,000                            & 40,000            & 34,144        & 85.36\%        & 10,000            & 651          & 6.51\%      \\
 Thunderbird & 10,000,000                       & 99,997                            & 79,997           & 837          & 1.05\%         & 20,000            & 29           & 0.15\%         \\
 \bottomrule
\end{tabular}}
\end{table*}

\subsection{Metrics}
We evaluate the performance of these methods using the widely adopted  $Precision$, $Recall$ and $F_1\!-\!score$. These metrics are calculated as follows:
\begin{equation}
    Precision=\frac{TP}{TP+FP}
\end{equation}
\begin{equation}
    Recall=\frac{TP}{TP+FN}
\end{equation}
\begin{equation}
    F_1\!-\!score =\frac{2 *Precision  * Recall }{ Precision + Recall }
\end{equation}
, where $TP$, $FN$, $FP$ represent true positives, false negatives and false positives, respectively. 

Precision refers to the percentage of correctly detected anomalies among all the anomalies identified by the model, while recall represents the percentage of anomalies that are correctly identified from all the real anomalies. The F$_1$-score combines these two metrics into a single measure, providing a balanced assessment of the model’s performance in detecting anomalies.

\subsection{Dataset}
\label{sec:data}
To evaluate our method for log-based anomaly detection, we selected four public datasets \cite{zhu2023loghub}: HDFS, BGL, Liberty, and Thunderbird. The details for each dataset are provided below:

\textbf{HDFS (Hadoop Distributed File System)}  dataset \cite{xu2009online} is generated by running Hadoop-based mapreduce jobs on over 200 Amazon EC2 nodes and contains a total of 11,175,629 log messages. These log messages are grouped into different log windows based on their \textit{block\_id}, which reflect program executions in the HDFS, resulting in 575,061 blocks. Among these, 16,838 blocks (2.93\%) indicate system anomalies.

\textbf{BGL (Blue Gene/L)} dataset  \cite{oliner2007supercomputers} is a supercomputing system log dataset collected from a BlueGene/L supercomputer system at Lawrence Livermore National Labs (LLNL). The dataset contains 4,747,963 log messages, each of which has been manually labeled as either normal or anomalous. There are 348,460 log messages (7.34\%) that are
labeled as anomalous.

\textbf{Thunderbird} dataset \cite{oliner2007supercomputers}  is a publicly accessible collection of log data sourced from the Thunderbird supercomputer at Sandia National Labs (SNL). This dataset consists of both normal and anomalous messages, each of which has been manually categorized. Although the dataset contains over 200 million log messages, we focus on a subset of 10 million continuous log messages for computational efficiency. This subset includes 4,937 anomalous log messages, representing approximately 0.049\% of the total.

\textbf{Liberty} dataset \cite{oliner2007supercomputers}  comprises system logs from the Liberty supercomputer at Sandia National Labs (SNL) in Albuquerque.  This supercomputer features 512 processors and 944 GB of memory, and the dataset contains over 200 million log messages. For computational efficiency, we sample 5 million consecutive log messages, among which 1,600,525 are identified as anomalous, constituting approximately 32.01\% of the total sampled messages.

In the context of HDFS, we adopt a session window strategy, which involves grouping log messages into sequences based on the \textit{block\_id} present in each log message. Each session is labeled using ground truth.
For other datasets, including BGL, Thunderbird, and Liberty, we utilize a sliding window strategy to group log messages, with a window size of 100 messages and a step size of 100  messages. A log sequence is deemed anomalous if it contains at least one anomalous log message according to the ground truth.  

Similar to existing work \cite{le2021log, le2022log, qi2023loggpt}, we split each dataset into a training set and a testing set with a ratio of 8:2 to evaluate the performance of a log-based anomaly detection approach.
For the HDFS dataset, we randomly split the log sequences into training and testing data.
In contrast, for the BGL, Thunderbird, and Liberty datasets, we adhere to a \textit{chronological split} \cite{le2022log} rather than a \textit{random split}.
The chronological split ensures that all log sequences in the training set precede those in the testing set, reflecting real-world conditions and mitigating potential data leakage from unstable log data.
Table \ref{tab:stastic} summarizes the statistics of the processed datasets used in the experiments.

\subsection{Performance Evaluation  (RQ1)}
\label{sec:exp_RQ1}
\begin{table*}[t]\renewcommand{\arraystretch}{}
\setlength{\tabcolsep}{2mm}{
\caption{Experimental results on HDFS, BGL, Liberty, and Thunderbird datasets. `Log parser' indicates whether a log parser is used by the method. The best results are highlighted in bold.}
\label{tab:res}
\begin{tabular}{ccccccccccccccc}
\toprule
\multirow{2}{*}{\diagbox {Methods}{Datasets} }  &   \multirow{2}{*}{Log parser} & \multicolumn{3}{c}{HDFS} &  \multicolumn{3}{c}{BGL}     &  \multicolumn{3}{c}{Liberty}    &  \multicolumn{3}{c}{Thunderbird} & \multirow{2}{*}{Avg. F$_1$}    \\ \cmidrule(lr){3-5} \cmidrule(lr){6-8} \cmidrule(lr){9-11} \cmidrule(lr){12-14}   &  & Prec. & Rec.  & F$_1$    & Prec. & Rec.  &  F$_1$    & Prec. & Rec.  &  F$_1$     & Prec. & Rec.  &  F$_1$   &  \\ \midrule
DeepLog    & $\usym{2714}$       & 0.835 & 0.994 & 0.908 & 0.166 & 0.988 & 0.285 & 0.751 & 0.855 & 0.800 & 0.017 & 0.966 & 0.033 & 0.506   \\
LogAnomaly & $\usym{2714}$      & 0.886 & 0.893 & 0.889 & 0.176 & 0.985 & 0.299 & 0.684 & 0.876 & 0.768 & 0.025 & 0.966 & 0.050 & 0.502 \\
PLELog     & $\usym{2714}$         & 0.893 & 0.979 & 0.934 & 0.595 & 0.880 & 0.710 & 0.795 & 0.874 & 0.832 & 0.808 & 0.724 & 0.764 & 0.810 \\
FastLogAD  & $\usym{2714}$     & 0.721 & 0.893 & 0.798 & 0.167 & \textbf{1.000} & 0.287 & 0.151 & \textbf{0.999} & 0.263 & 0.008 & 0.931 & 0.017 & 0.341  \\
LogBERT    & $\usym{2714}$      & 0.989 & 0.614 & 0.758 & 0.165 & 0.989 & 0.283 & 0.902 & 0.633 & 0.744 & 0.022 & 0.172 & 0.039 & 0.456 \\ 
LogRobust  & $\usym{2714}$       & 0.961 & \textbf{1.000} & 0.980 & 0.696 & 0.968 & 0.810 & 0.695 & 0.979 & 0.813 & 0.318 & \textbf{1.000} & 0.482 & 0.771 \\
CNN        & $\usym{2714}$        & 0.966 & \textbf{1.000} & 0.982 & 0.698 & 0.965 & 0.810 & 0.580 & 0.914 & 0.709 & 0.870 & 0.690 & 0.769  & 0.818 \\
LogLLaMA        & $\usym{2714}$  & 0.987 & 0.950 & 0.968 & 0.196 & 0.983 & 0.326 & 0.785 & 0.920 & 0.847 & 0.068 & 0.966 & 0.127 & 0.567 \\ 
EagerLog     & $\usym{2714}$  & 0.928 & 0.935 & 0.931 & 0.469 & \textbf{1.000} & 0.639 & 0.575 & 0.969 & 0.722 & 0.296 & \textbf{1.000} & 0.457 & 0.687 \\
NeuralLog & $\usym{2717}$  & 0.971 & 0.988 & 0.979 & 0.792 & 0.884 & 0.835 & 0.875 & 0.926 & 0.900 & 0.794 & 0.931 & 0.857 & 0.893 \\
RAPID      & $\usym{2717}$      & \textbf{1.000} & 0.859 & 0.924 & \textbf{0.874} & 0.399 & 0.548 & 0.911 & 0.611 & 0.732 & 0.200 & 0.207 & 0.203 & 0.602  \\
BGE-Trans &  $\usym{2717}$ & 0.975 & 0.988 & 0.981 & 0.799 & 0.848 & 0.823 & 0.894 & 0.896 & 0.895 & 0.839 & 0.897 & 0.867 & 0.891 \\ 
LogLLM    &  $\usym{2717}$  & 0.994 & \textbf{1.000} & \textbf{0.997} &  0.861 & 0.979 & \textbf{0.916} & \textbf{0.992} & 0.926 & \textbf{0.958} & \textbf{0.966} & 0.966 & \textbf{0.966} & \textbf{0.959}  \\ \bottomrule
\end{tabular}}
\end{table*}

Table \ref{tab:res} presents the experimental results of various log-based anomaly detection methods on the HDFS, BGL, Liberty, and Thunderbird datasets. The best results are highlighted in bold. We have the following observations:

The proposed LogLLM achieves the highest F$_1$-score across all datasets. On average, LogLLM's F$_1$-scores are 6.6\% better than the best existing method, NeuralLog, demonstrating its effectiveness in log-based anomaly detection.  
Despite the adoption of LLMs in FastLogAD, LogBERT, NeuralLog, EagerLog, LogLLaMA, RAPID and BGE-Trans for anomaly detection, their performance remains unsatisfactory. 
FastLogAD and LogBERT leverage BERT, a Transformer encoder-based LLM, to detect anomalies via log sequence reconstruction errors, while LogLLaMA employs Llama, a Transformer decoder-based LLM, to predict the next template ID for anomaly detection; in all cases, the input consists of sequences of log template IDs extracted from log messages using log parsers, which lack semantic information.
During testing, these methods have no knowledge of unseen template IDs, i.e., template IDs that do not appear in the training dataset.
Consequently, they cannot correctly reconstruct or predict these unseen template IDs, causing log sequences containing them to be frequently misclassified as anomalies, which increases false positives and reduces precision.
This issue is particularly pronounced in certain benchmark datasets, where the proportion of unseen template IDs in the test dataset, defined as $\frac{\#\text{unseen template IDs}}{\#\text{total template IDs}}$, is 41.1\% in BGL, 72.9\% in Liberty, and  83.3\% in Thunderbird. 
In contrast, NeuralLog, RAPID and BGE-Trans utilize Transformer encoder-based LLMs to extract semantic vectors from log messages. 
However, NeuralLog and BGE-Trans rely on relatively lightweight classification models, and their suboptimal performance indicates that simply combining a general-purpose encoder with a lightweight classifier is insufficient. 
RAPID, meanwhile, performs anomaly sequence classification through distance-based comparison, which lacks the capacity to explicitly model complex sequential dependencies among log messages.
Although EagerLog also employs a Transformer decoder-based LLM, it performs anomaly detection through retrieval-augmented generation. Given the diverse semantic and sequential patterns of log anomalies, a limited number of retrieved log sequences cannot adequately represent all anomaly scenarios. 
Consequently, its performance heavily depends on the quality and representativeness of the retrieved examples, limiting its ability. 
In contrast, LogLLM leverages both Transformer encoder-based and decoder-based LLMs and is trained on the target dataset for anomaly detection, fully harnessing semantic understanding, sequence modeling, and the rich prior knowledge embedded in LLMs, leading to superior anomaly detection performance.

\begin{table}[]\renewcommand{\arraystretch}{}
\setlength{\tabcolsep}{1.6mm}{
\caption{Runtime comparison.}
\label{tab:Running_time}
\begin{tabular}{cccc}
\toprule
           & Train (min) & Test (min) & Inference Latency (ms/sample) \\ \midrule
DeepLog           & 72.17   & 3.42 &  7.37\\
LogAnomaly        & 156.16  & 7.25 &   17.68 \\
PLELog            & 315.47  & 33.59 &  82.17\\
FastLogAD         & 254.17  & 0.29 & 0.87 \\
LogBERT           & 429.04  & 43.77 & 93.42 \\
LogRobust         & 108.42  & 2.48 &  5.90\\
CNN               & 98.16   & 2.16 & 4.91\\
LogLLaMA          & 3240.45 & 59.15 & 124.21 \\    
EagerLog          & 59.32   & 82.71 &  172.84 \\
NeuralLog         & 267.46  & 21.44 &  52.31 \\
RAPID             & 63.98   & 38.43 &  92.43 \\
BGE-Trans       & 241.91  & 19.80 &  49.75 \\
LogLLM         & 1,065.15 & 64.48  & 150.06\\ \bottomrule
\end{tabular}
}
\end{table}
Moreover, LogLLM achieves a balance between precision and recall, indicating that it maintains low false alarm rates and minimizes missed reports.
In contrast, methods like FastLogAD are excessively sensitive to anomalies, often resulting in numerous false alarms. For example, on the BGL dataset, despite FastLogAD having a recall of 1, it only achieves a precision of 0.167, making it impractical for real-world use. Similarly, methods such as DeepLog, LogAnomaly, LogBERT and LogLLaMA exhibit similar issues.
On the other hand, RAPID is not sensitive enough to anomalies, leading to many undetected anomalies. For instance, on the BGL dataset, RAPID achieves a precision of 0.874 but a recall of only 0.399.

In summary, our experimental datasets generated through chronological splitting often produce training sets that fail to cover all log sequence patterns present in the test set, making semantic-aware anomaly detection and strong model generalization essential. 
Existing methods struggle to address this challenge. LogLLM, however, leverages LLMs with rich knowledge to effectively detect anomalies in unseen log templates and sequential patterns from a semantic perspective, making it a highly effective solution for log-based anomaly detection.

\textbf{Effect of labeled anomalies}.
As illustrated in Table \ref{tab:res}, in contrast to methods such as DeepLog, LogAnomaly, FastLogAD, LogBERT, EagerLog, RAPID and LogLLaMA, which require clean datasets devoid of anomalies to build anomaly detection models, methods like PLELog, LogRobust, CNN, NeuralLog, BGE-Trans and LogLLM demonstrate superior performance. These models are trained using not only normal samples but also labeled anomalies.
For instance, these six methods achieve an average F$_1$-score above 0.771 across four datasets, whereas others that do not utilize labeled anomalies perform poorly, with an average F$_1$-score below 0.687  across four datasets.
This demonstrates that incorporating labeled anomalies can provide a significant advantage to anomaly detection methods.

\begin{table*}[t]\renewcommand{\arraystretch}{}
\setlength{\tabcolsep}{2.84mm}{
\caption{Effects of different preprocessing techniques on HDFS, BGL, Liberty, and Thunderbird datasets. The best results are highlighted in bold.}
\label{tab:effectpre}
\begin{tabular}{cccccccccccccc}
\toprule
      & \multicolumn{3}{c}{HDFS} & \multicolumn{3}{c}{BGL} & \multicolumn{3}{c}{Liberty} & \multicolumn{3}{c}{Thunderbird} & \multirow{2}{*}{Avg. F$_1$} \\ \cmidrule(lr){2-4} \cmidrule(lr){5-7} \cmidrule(lr){8-10} \cmidrule(lr){11-13}  
                  & Prec.  & Rec.   & F$_1$     & Prec.  & Rec.   & F$_1$    & Prec.   & Rec.    & F$_1$     & Prec.     & Rec.     & F$_1$       &                          \\ \midrule
Raw               & 0.994  & 0.991  & 0.993  & \textbf{0.943}  & 0.767  & 0.846 & 0.911   & 0.908   & 0.909   & 0.806     & 0.862    & 0.833    & 0.895                    \\
Template ID    & \textbf{0.995}  & 0.945  & 0.969  & 0.775  & 0.286  & 0.418 & \textbf{0.994}   & 0.270   & 0.425   & \textbf{1.000}     & 0.379    & 0.550    & 0.591                    \\
Template  & 0.991 & \textbf{1.000} & 0.995 & 0.861 & 0.919 & 0.889 & 0.968 & \textbf{0.931} & 0.949 & 0.950 & 0.655 & 0.776 & 0.902 \\         
RE (LogLLM) & 0.994 & \textbf{1.000} & \textbf{0.997} &  0.861 & \textbf{0.979} & \textbf{0.916} & 0.992 & 0.926 & \textbf{0.958} &  0.966  & \textbf{0.966} & \textbf{0.966} & \textbf{0.959}     
 \\ \bottomrule
\end{tabular}}
\end{table*}

\textbf{Runtime comparison}.
The time consumption of each method is presented in Table \ref{tab:Running_time}. These results have been averaged across all the datasets.
Although EagerLog and RAPID do not require training a deep learning model, they still need to construct embedding vectors for the training dataset to support retrieval.
FastLogAD shows a relatively high training cost (254.17 min) but the shortest testing time (0.29 min) because only the discriminator is used during inference.
As expected, methods that rely on LLMs incur substantially higher computation. 
In particular, LogLLaMA has the longest training time (3240.45 min), which is roughly 3.0× that of LogLLM (1065.15 min) and about 7.6× that of LogBERT (429.04 min). 
This extreme training cost is attributable to its two-stage pipeline: a large-scale pretraining stage followed by a RL stage; RL-style fine-tuning typically requires many additional optimization steps and environment rollouts, resulting in considerably higher training overhead than supervised fine-tuning alone.
Although our proposed LogLLM also incurs relatively high training overhead, its training time remains lower than LogLLaMA’s, and its testing time remains acceptable when compared to other methods that utilize LLMs, such as LogBERT,  RAPID, EagerLog and LogLLaMA. 
The slower inference of LogLLM compared to LogLLaMA is mainly attributed to its autoregressive text generation process, in which output tokens are generated sequentially rather than through a single-step prediction as in LogLLaMA. 
Although LogLLM requires substantial training time, training is performed offline. During online detection, LogLLM’s inference latency is markedly shorter than the corresponding sample-generation interval across all datasets: 0.079s versus 109.62s on HDFS, 0.142s versus 393.51s on BGL, 0.249s versus 46.49s on Liberty, and 0.131s versus 7.51s on Thunderbird. Therefore, LogLLM can complete inference for each sample before the next sample is generated, making it suitable for scenarios requiring near-real-time detection.

\subsection{Different Preprocessing Techniques (RQ2)}
\label{sec:effectpre}   
We evaluate the effectiveness of the different preprocessing techniques. The results are shown in Table \ref{tab:effectpre}. In this table, `Raw' indicates that the content of log messages is not preprocessed and is directly input into the proposed deep model. `Template' indicates that sequences of log templates produced by Drain \cite{he2017drain}, a log parser, are used as input for the proposed deep model. `Template ID' signifies that the IDs of log templates, obtained by Drain, are simply encoded into numeric vectors using an embedding layer instead of LLM$_\mathcal{E}$. The preprocessing technique `Template ID' renders the model unable to capture the semantic information within log messages.
`RE' indicates that regular expressions, as introduced in Section \ref{sec:Preprocessing}, are used for preprocessing log messages.

As anticipated, the preprocessing technique `RE' yields the highest F$_1$-score across all datasets. Conversely, the preprocessing technique `Template ID' consistently results in the lowest F$_1$-score across all datasets, averaging 36.8\% lower than that of `RE'.
This can be attributed to the fact that `Template ID' carries no semantic information, thereby limiting the model’s ability to detect anomalies from a natural language perspective. Moreover, when encountering templates present in the test dataset but absent in the training dataset, the model lacks prior knowledge of these IDs and cannot conduct anomaly detection correctly.  
This issue is particularly pronounced in certain benchmark datasets, where the proportion of unseen template IDs in the test dataset, defined as $\frac{\#\text{unseen template IDs}}{\#\text{total template IDs}}$, is 41.1\% in BGL, 72.9\% in Liberty, and  83.3\% in Thunderbird.
The preprocessing techniques `Raw' and `Template' result in relatively good performance, but their F$_1$-scores are still 6.4\% and 5.7\% lower than that of `RE', respectively. For the preprocessing technique `Raw', the variable parts (parameters that carry dynamic runtime information) within the content of each log message have little influence on anomaly detection. However, due to their high randomness, they can confuse the model, making it difficult to discern anomalies.
For the preprocessing technique `Template', although it can provide semantic information, the log parser is not always reliable and may sometimes discard informative constant parts, thereby causing information loss, or retain irrelevant variable parts, thereby introducing unnecessary variations, which may ultimately degrade the model's ability to identify anomalies.

\begin{table*}[t]\renewcommand{\arraystretch}{}
\setlength{\tabcolsep}{1mm}{
\caption{Effects of the embedder (LLM$_\mathcal{E}$ \& projector) and model size of LLM$_\mathcal{D}$, where `Mem.' indicates GPU memory usage (GB), and `Tim.' indicates training time (Minutes). `-' indicates an out-of-memory (OOM) error.}
\label{tab:llms}
   \resizebox{\linewidth}{!}{
\begin{tabular}{ccccccccccccccccccccc}
\toprule
                     & \multicolumn{5}{c}{HDFS}        & \multicolumn{5}{c}{BGL}         & \multicolumn{5}{c}{Liberty}     & \multicolumn{5}{c}{Thunderbird} \\
                      \cmidrule(lr){2-6}  \cmidrule(lr){7-11}  \cmidrule(lr){12-16}  \cmidrule(lr){17-21}
                         & Prec. & Rec.  & F$_1$    & Mem. & Tim. & Prec. & Rec.  & F$_1$    & Mem. & Tim. & Prec. & Rec.  & F$_1$    & Mem. & Tim. & Prec. & Rec.  & F$_1$    & Mem. & Tim. \\ \midrule
L.-1B           & 0.986                     & 0.995                    & 0.991                  & 16.5                        & 2022.1                    & -                        & -                      & -                           & -                         & -                        & 0.960                  & 0.699                       & 0.809                     & 42.6                     & 443.2                  & \textbf{1.000}                       & 0.724 & 0.840 & 44.5 & 1732.1 \\

Emb. \& L.-1B & \textbf{0.996}                     & 0.996                    & 0.996                  & 8.0                         & 1412.2                    & 0.734                    & 0.944                  & 0.825                       & 32.4                      & 187.1                    & 0.950                  & 0.905                       & 0.927                     & 29.3                     & 173.2                  & 0.875                       & 0.966 & 0.918 & 32.4 & 715.1  \\

L.-8B             & 0.988                     & 0.997                    & 0.992                  & 43.0                        & 4712.1                    & -                        & -                      & -                           & -                         & -                        & -                      & -                           & -                         & -                        & -                      & -                           & -     & -     & -    & -     \\

Emb. \& L.-8B   & 0.994                     & \textbf{1.000}                    & \textbf{0.997}                  & 16.6                        & 2168.2                    & \textbf{0.861}                    & \textbf{0.979}                  & \textbf{0.916}                       & 38.0                      & 396.2                    & \textbf{0.992}                  & \textbf{0.926}                       & \textbf{0.958}                     & 36.1                     & 412.1                  & 0.966                       & \textbf{0.966} & \textbf{0.966} & 38.2 & 1284.2 
 \\ \bottomrule
\end{tabular}}
}
\end{table*}

\subsection{Effect of the Embedder (RQ3)}
\label{sec:ch}
We investigate how the embedder (LLM$_\mathcal{E}$ and projector) influences the performance of LogLLM. The results are presented in Table \ref{tab:llms}. In this table, `L.-1B' and `L.-8B' refer to directly inputting the log sequence, by concatenating log messages with semicolons (;) as separators into a long string, into  \textit{Llama-3.2-1B} model\footnote{https://huggingface.co/meta-llama/Llama-3.2-1B} and \textit{Llama-3-8B} model, respectively. `Emb. \& L.-1B' and `Emb. \& L.-8B'  represents LogLLM configured with `Llama-3.2-1B' and `Llama-3-8B' serving as LLM$_\mathcal{D}$. We have the following observations:

Incorporating the embedder significantly reduces GPU memory consumption and enhances detection performance. For example, on HDFS, `Emb. \& L.-1B' achieves an F$_1$-score of 0.996 with only 8.0 GB of memory, compared to 0.991 and 16.5 GB for its embedder-free counterpart `L.-1B'. Similarly, on BGL, direct input variants (`L.-1B' and `L.-8B') fail due to OOM errors, whereas `Emb. \& L.-8B' successfully runs with 38.0 GB of memory and achieves an F$_1$-score of 0.916.
These results demonstrate that the embedder effectively compresses the semantic information of each log message into a single semantic vector, significantly reducing the input length and memory consumption compared with directly tokenizing raw log message sequences, thereby preventing OOM issues caused by excessively long log sequences.
Moreover, this semantic vector-based representation provides clearer message boundaries within log sequences, allowing the Transformer decoder-based LLM to directly capture sequential dependencies among log messages instead of the token-level fragmentation introduced by direct tokenization. Therefore, the proposed embedding-based representation improves both computational efficiency and sequence modeling capability for long log sequences.
Despite the additional embedder, overall training time decreases. For instance, `Emb. \& L.-1B' reduces training time on Thunderbird from 1732.1 minutes to 715.1 minutes. This reduction arises because the majority of the time cost comes from LLM$_\mathcal{D}$. By using the embedder to transform each log message into a single embedding vector, the input length for LLM$_\mathcal{D}$ is shortened, effectively accelerating training.

In summary, the embedder is essential for LogLLM, serving as both a semantic abstraction layer and a computationally efficient intermediary between log sequences and the generative decoder, which reduces GPU memory consumption, improves performance, and accelerates training.


\subsection{Effect of the LLM$_\mathcal{D}$ (RQ4)}
To investigate how different LLM$_\mathcal{D}$ configurations influence performance, we evaluate LogLLM from three perspectives: model size, model architecture, and comparison with lightweight classifier alternatives. 


\textbf{Impact of model size}.
To examine the influence of model size, we evaluate Transformer decoder-based LLMs from the Llama family, which share the same architecture but differ in parameter scale, as shown in Table \ref{tab:llms}. 
The results indicate that larger Llama models consistently yield better detection performance, at the expense of higher GPU memory consumption and longer training time.
Specifically, on average, compared to using Llama-3.2-1B, adopting Llama-3-8B improves the F$_1$-score by 4.3\%, but increases GPU memory usage by 6.7 GB and extends training time by 443.3 minutes.
This improvement can be attributed to the stronger representation capacity of larger models, which enables them to capture more subtle dependencies among log messages and better model the complex sequential patterns underlying normal and anomalous behaviors.
However, the substantial computational overhead suggests a diminishing return when scaling up model size.
In practice, this trade-off implies that smaller models may offer a more favorable balance between accuracy and efficiency for real-time or resource-constrained log analysis scenarios, whereas larger models are more suitable for offline analysis tasks where accuracy is prioritized over latency.

\begin{table*}[t]\renewcommand{\arraystretch}{}
\setlength{\tabcolsep}{2.58mm}{
\caption{Performance comparison of LLM$_\mathcal{D}$ architectures and lightweight classifier alternatives.}
\label{tab:effectma}
\begin{tabular}{cccccccccccccc}
\toprule
      & \multicolumn{3}{c}{HDFS} & \multicolumn{3}{c}{BGL} & \multicolumn{3}{c}{Liberty} & \multicolumn{3}{c}{Thunderbird} & \multirow{2}{*}{Avg. F$_1$} \\ \cmidrule(lr){2-4} \cmidrule(lr){5-7} \cmidrule(lr){8-10} \cmidrule(lr){11-13}  
                  & Prec.  & Rec.   & F$_1$     & Prec.  & Rec.   & F$_1$    & Prec.   & Rec.    & F$_1$     & Prec.     & Rec.     & F$_1$       &                          \\ \midrule
Llama-3-8B & 0.994 & 1.000 & 0.997 & 0.861 & 0.979 & 0.916 & 0.992 & 0.926 & 0.958 & 0.966 & 0.966 & 0.966 & 0.959 \\
Qwen3-8B   & 0.997 & 1.000 & 0.999 & 0.927 & 0.968 & 0.947 & 0.943 & 0.988 & 0.965 & 0.903 & 0.966 & 0.933 & 0.961 \\
GLM-4-9B   & 0.991 & 1.000 & 0.996 & 0.908 & 0.935 & 0.922 & 0.940 & 0.969 & 0.955 & 0.875 & 0.966 & 0.918 & 0.947 \\
MPT-7B     & 0.994 & 1.000 & 0.997 & 0.880 & 0.919 & 0.899 & 0.951 & 0.985 & 0.968 & 0.966 & 0.966 & 0.966 & 0.957 \\
\hdashline
MLP                 & 0.947 & 0.971 & 0.959 & 0.668 & 0.764 & 0.713 & 0.836 & 0.862 & 0.849 & 0.676 & 0.793 & 0.730 & 0.813 \\
Transformer Encoder & 0.969 & 0.987 & 0.978 & 0.778 & 0.886 & 0.828 & 0.875 & 0.912 & 0.893 & 0.800 & 0.966 & 0.875 & 0.894  \\
Transformer Decoder & 0.970 & 0.997 & 0.983 & 0.794 & 0.851 & 0.822 & 0.856 & 0.912 & 0.883 & 0.839 & 0.897 & 0.867 & 0.889  \\
\bottomrule
\end{tabular}}
\end{table*}

\textbf{Impact of model architectures}.
We evaluate LogLLM using different LLM$_\mathcal{D}$ architectures of comparable size, including \textit{Llama-3-8B} model, \textit{Qwen3-8B} model,
\textit{GLM-4-9B} model\footnote{https://huggingface.co/zai-org/GLM-4-9B-0414}
and \textit{MPT-7B} model\footnote{https://huggingface.co/mosaicml/mpt-7b}. As shown in Table~\ref{tab:effectma}, all models achieve high F$_1$-scores across the HDFS, BGL, Liberty, and Thunderbird datasets. The Qwen3-8B model achieves the highest average F$_1$-score of 0.961, while the GLM-4-9B model attains the lowest average F$_1$-score of 0.947. The maximum difference in average F$1$-score is only 0.014, demonstrating that LogLLM is largely insensitive to the choice of LLM$_\mathcal{D}$ architecture when model sizes are similar. 

\begin{table*}[t]\renewcommand{\arraystretch}{}
\setlength{\tabcolsep}{2.5mm}{
\caption{Ablation study of training stages and comparison of minority-class balancing strategies on HDFS, BGL, Liberty, and Thunderbird datasets,  where `\textit{w/o}' denotes `\textit{without}' and `\textit{w/}' denotes `\textit{with}'.  The best results are highlighted in bold.}
\label{tab:Ablation_Study}
\begin{tabular}{cccccccccccccc}
\toprule
     & \multicolumn{3}{c}{HDFS} & \multicolumn{3}{c}{BGL} & \multicolumn{3}{c}{Liberty} & \multicolumn{3}{c}{Thunderbird} & \multirow{2}{*}{Avg. F$_1$} \\ \cmidrule(lr){2-4} \cmidrule(lr){5-7} \cmidrule(lr){8-10} \cmidrule(lr){11-13}  
                  & Prec.  & Rec.   & F$_1$     & Prec.  & Rec.   & F$_1$    & Prec.   & Rec.    & F$_1$      & Prec.     & Rec.     & F$_1$       &                          \\ \midrule
w/o Stage 1           & 0.991                     & \textbf{1.000}                    & 0.995                  & 0.578                     & 0.971                    & 0.725                  & 0.685                     & 0.290                    & 0.408                  & 0.381                     & 0.828                    & 0.522                  & 0.662                                        \\
w/o Stage 2           & 0.994                     & \textbf{1.000}                    & \textbf{0.997}                  & 0.858 &	0.920  & 0.888 
                  & 0.995                     & 0.906                    & 0.949                  & 0.848                     & 0.966                    & 0.903                  & 0.934                                     \\
w/o Stage 1\&2        & 0.992                     & \textbf{1.000}                    & 0.996                  &  0.853 	& 0.882 &	0.868 
                & 0.995                     & 0.906                    & 0.949                  & 0.897                     & 0.897                    & 0.897                  & 0.927                                       \\
w/o Stage 3           & 0.993                     & 0.999                    & 0.996                  & 0.704                     & 0.776                    & 0.738                  & \textbf{1.000}                     & 0.684                    & 0.812                  & 0.958                     & 0.793                    & 0.868                  & 0.854                                        \\
w/o MCO \& w/ CSL   &    0.993 & \textbf{1.000} & 0.996 & 0.842 & 0.936 & 0.886 & 0.985 & \textbf{0.928} & 0.956 & 0.486 & 0.586 & 0.531 & 0.842  \\
LogLLM         & \textbf{0.994} & \textbf{1.000} & \textbf{0.997} &  \textbf{0.861} & \textbf{0.979} & \textbf{0.916} & 0.992 & 0.926 & \textbf{0.958} & \textbf{0.966} & \textbf{0.966} & \textbf{0.966} & \textbf{0.959}
 \\ \bottomrule
\end{tabular}}
\end{table*}

\textbf{Comparison with lightweight classifiers}.
To provide controlled comparisons with simpler alternatives, we replace LLM$_\mathcal{D}$ with several lightweight classifier architectures built on top of the same LLM$_\mathcal{E}$. Specifically, we compare LogLLM with three variants, as shown in Table~\ref{tab:effectma}. `MLP' denotes a model that concatenates the semantic vectors of all log messages output by LLM$_\mathcal{E}$ and feeds the resulting representation into a two-layer MLP for binary classification.
`Transformer Encoder' denotes a model that feeds the semantic vectors output by LLM$_\mathcal{E}$ into a two-layer Transformer encoder with positional encoding, followed by mean pooling and a linear layer for binary classification.  
`Transformer Decoder'  denotes a model that feeds the semantic vectors output by LLM$_\mathcal{E}$ into a two-layer Transformer decoder with positional encoding, and then passes the final hidden state of the last token to a linear layer for binary classification.
The results show that although these lightweight classifiers are simpler and more computationally efficient, they consistently perform worse than LogLLM. 
These results suggest that a larger pretrained Transformer decoder-based LLM can be beneficial for log anomaly detection.

\subsection{Effect of the Training Procedure (RQ5)}
\label{sec:exptp}
To investigate how different training procedures influence the performance of LogLLM, we first conduct an ablation study to analyze the contribution of each training stage. Furthermore, we perform a sensitivity analysis on the impact of the number of samples used in Stage 1.

\textbf{Ablation study of different training stages}.
We investigate the effect of each training procedure through an ablation study. The results are presented in Table \ref{tab:Ablation_Study}. We have the following observations:
Skipping any training stage results in a decrease in the F$_1$-score across all datasets, demonstrating the effectiveness of our three-stage training procedure.
It is noteworthy that training without stage 1 leads to the worst performance, with the F$_1$-score averaged across all datasets decreasing by as much as 29.7\%. However, training without stages 1\&2 (only adopting training stage 3: fine-tuning the entire model) yields acceptable performance, with only a 3.2\% decrease in the average F$_1$-score. 
This demonstrates that fine-tuning LLM$_\mathcal{D}$ for response format adaptation (Stage 1) is essential before training the embedder (LLM$_\mathcal{E}$ and projector) of log messages (Stage 2). 
Without Stage 1 (i.e., directly training the embedder), the embedder may be misdirected, resulting in incorrect semantic capture of log messages and model failure, because the initial output distribution of the LLM$_\mathcal{D}$ does not match the expected templates, causing the embedder to receive misleading gradient signals during early training.
Training without stage 3 yields relatively poor performance, with an average F$_1$-score decrease of 10.5\%. This indicates that sequentially fine-tuning LLM$_\mathcal{D}$ and training the embedder alone is insufficient for the model to capture anomalous patterns; cohesive fine-tuning of the entire model is essential.
Training without stages 2 and 1\&2 also results in a performance decrease, with average F$_1$-score reductions of 2.5\% and 3.2\%, respectively. This demonstrates that individually training the embedder  before fine-tuning the entire model can also enhance performance. This stage allows the embedder to generate better semantic embedding vectors of log messages for LLM$_\mathcal{D}$ to discern anomalies.
In summary, our proposed three-stage training procedure is well-suited for our deep model in log-based anomaly detection.

\begin{figure}[tb]
	\centering
    \includegraphics[scale=0.35]{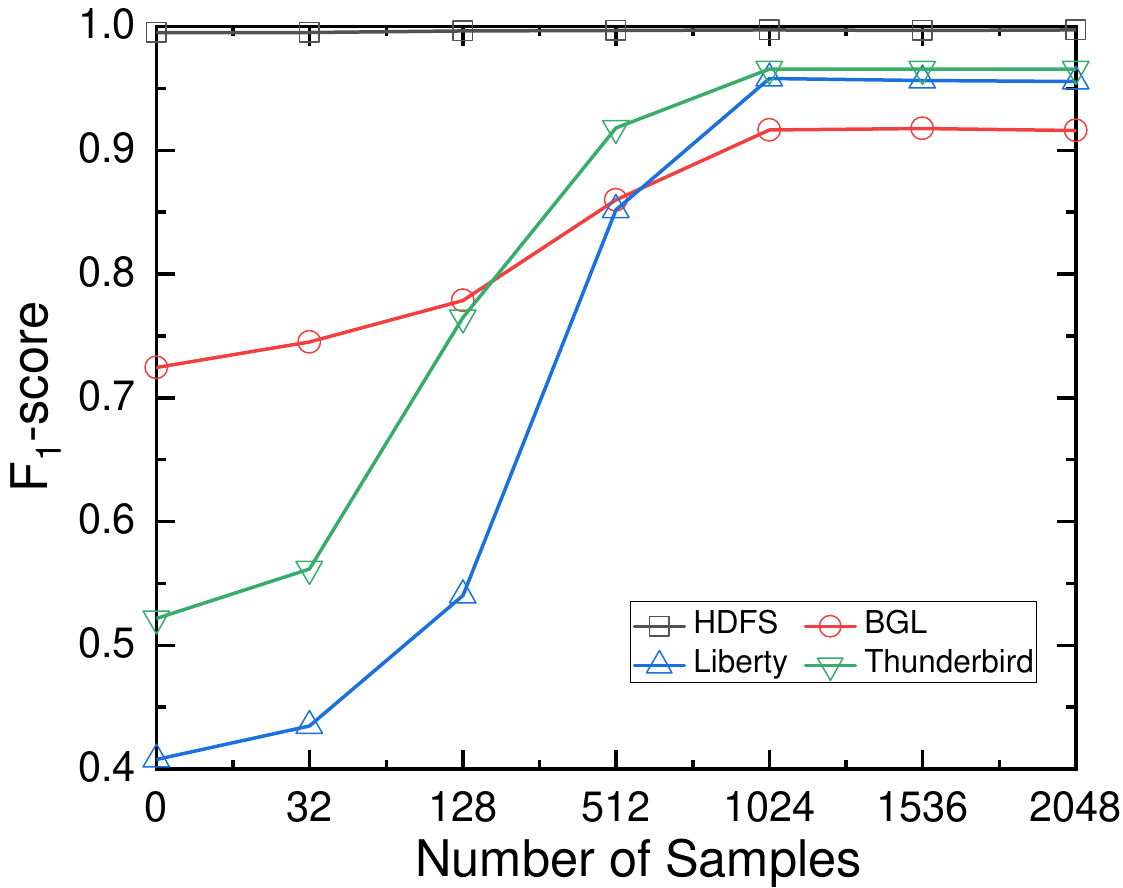}
 	\caption{Sensitivity analysis of the number of samples used in Stage 1 on four datasets.}
	\label{fig:SensitivityS1}
\end{figure}

\textbf{Sensitivity analysis of the number of samples used in Stage 1}.
To investigate the impact of the number of samples used in Stage 1, we vary the number of samples and evaluate the performance of LogLLM on four datasets. As shown in Fig. \ref{fig:SensitivityS1}, the performance generally improves with the increase in the number of samples, indicating that more samples provide richer information for learning the desired response format.
However, the performance gradually becomes stable when the number of samples exceeds 1,024. Specifically, all four datasets achieve stable performance with more than 1,024 samples, indicating that Stage 1 does not require a large number of samples. Therefore, approximately 1,000 samples are sufficient to achieve near-optimal anomaly detection performance across different datasets.

\begin{figure*}[t]
    	\centering
    \subfloat[Precision]{
	\centering
	\includegraphics[scale=0.35]{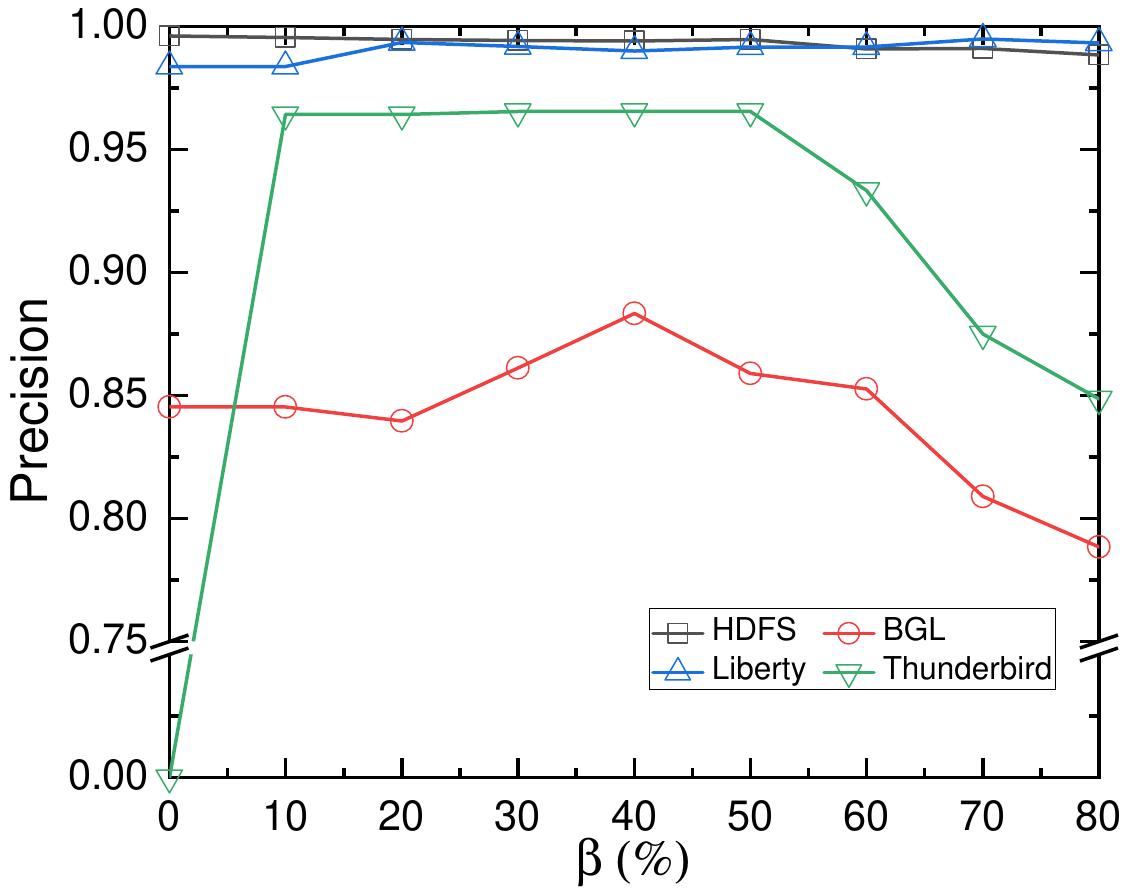}
	\label{fig:PrecisionG}
	}
	\hfil
\subfloat[Recall]{
	\centering
	\includegraphics[scale=0.35]{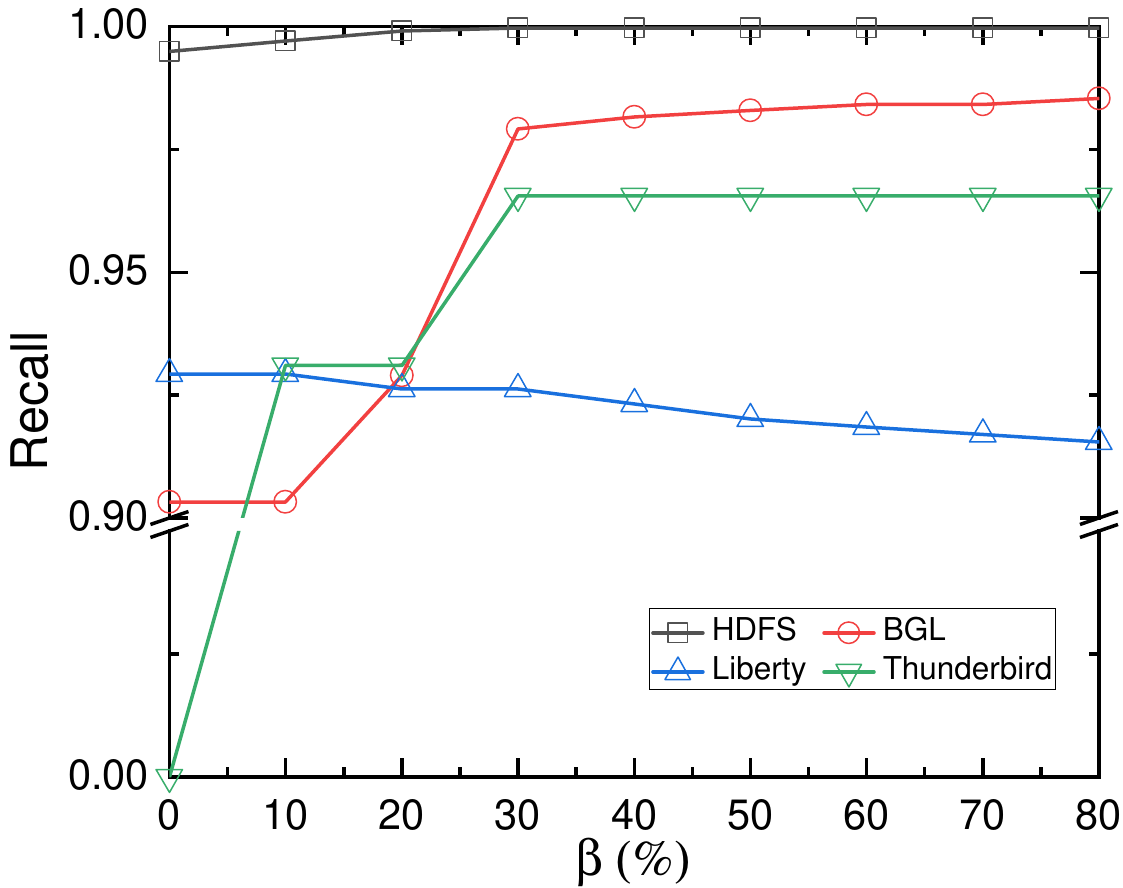}
	\label{fig:RecallG}
	}\\
\subfloat[F$_1$-score]{
	\centering
	\includegraphics[scale=0.35]{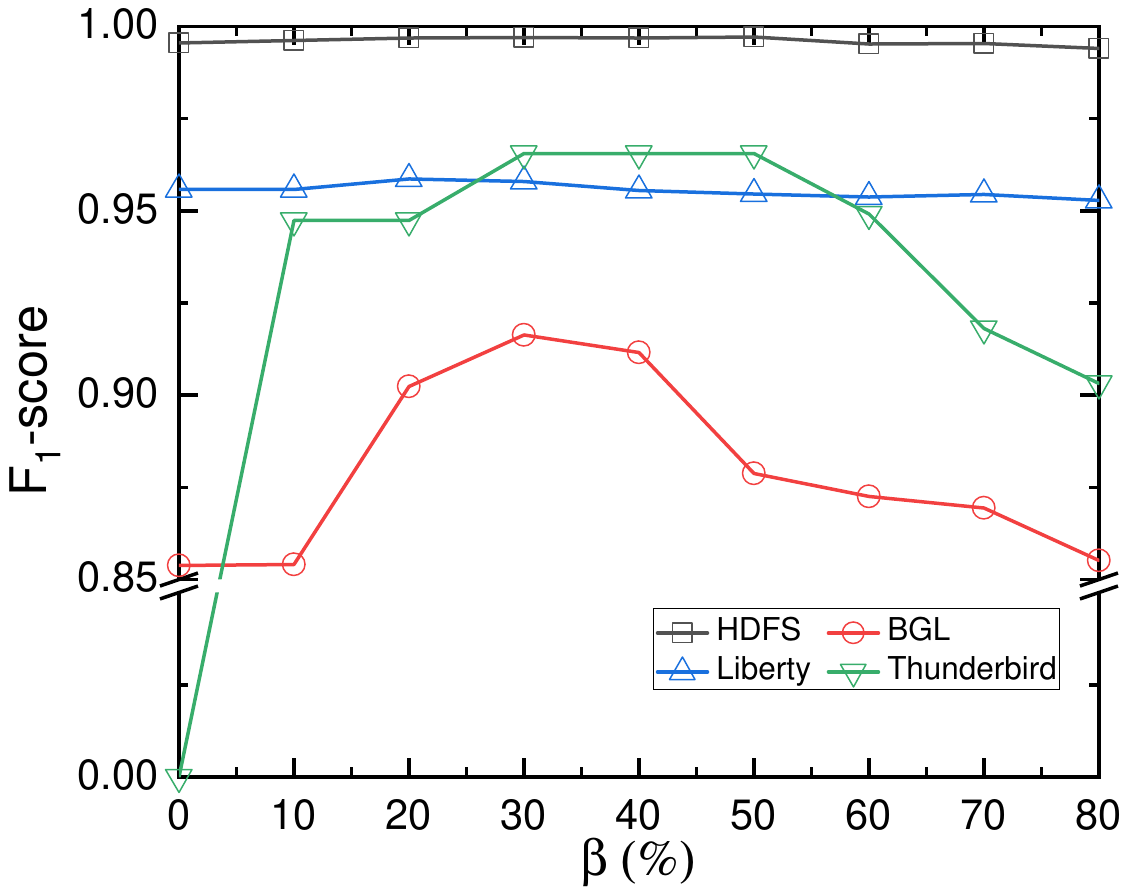}
	\label{fig:FG}
	}\hfil
 \subfloat[Training time]{
	\centering
	\includegraphics[scale=0.35]{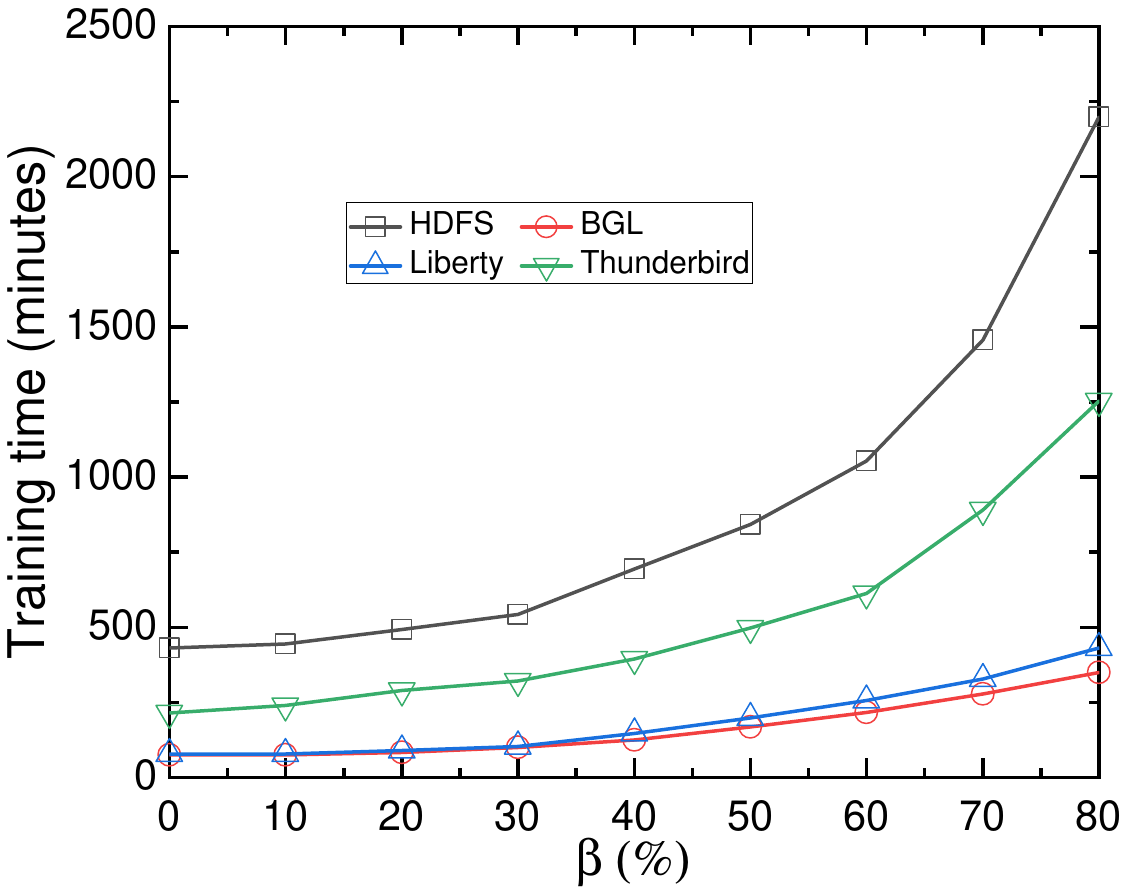}
	\label{fig:tt}
	}
	\caption{Impact of minority class oversampling.}
 \label{fig:beta}
\end{figure*}

\subsection{Effect of Minority Class Oversampling (RQ6)}
Note that normal and anomalous samples in the training datasets are imbalanced, as shown in Table \ref{tab:stastic}. For the HDFS, BGL, and Thunderbird datasets, normal samples outnumber anomalous samples, while in the Liberty dataset, anomalous samples exceed normal samples. Therefore, the minority class differs across datasets. 
To investigate the effectiveness of minority class oversampling in improving LogLLM under imbalanced anomaly detection scenarios, we first compare our oversampling strategy with a class-weighted cost-sensitive learning approach. Furthermore, we analyze the impact of different oversampling levels controlled by the hyperparameter $\beta$.

\textbf{Comparison with class-weighted cost-sensitive learning}.
To evaluate whether the improvement brought by oversampling can be achieved by adjusting the loss function, we replace our minority class oversampling strategy with a class-weighted Cost-Sensitive Learning (CSL) strategy (`w/o MCO \& w/ CSL'). 
Specifically, we assign a larger weight to the minority class during optimization to increase its contribution to the loss. In our experiments, the default oversampling setting uses $\beta=0.3$, which adjusts the minority-to-majority class ratio to 3:7. To ensure a fair comparison, we set the class weights in cost-sensitive learning such that the effective contribution ratio between the minority and majority classes is consistent with the oversampling strategy:
$\frac{N_{min} w_{min}}{N_{maj} w_{maj}}=\frac{3}{7}$, where $N_{min}$ and $N_{maj}$ denote the numbers of minority and majority class samples, respectively, and $w_{min}$ and $w_{maj}$ denote the weights assigned to the minority and majority classes in the loss function. We set $w_{maj}=1$ and calculate $w_{min}$ accordingly for each dataset.
Table \ref{tab:Ablation_Study} reports the results. 
Compared with LogLLM, the variant using CSL achieves lower F$_1$-score across all datasets, especially on the highly imbalanced Thunderbird dataset. 
Although CSL improves anomaly detection performance by increasing the contribution of the minority class during optimization, it cannot fully match the performance of minority class oversampling. 
This difference can be attributed to their different effects on the training process. 
Minority class oversampling increases the sampling frequency of minority-class samples during training, allowing the model to observe minority-class patterns more frequently. 
In contrast, CSL only adjusts the loss contribution of existing minority-class samples while maintaining the original sampling distribution. 
Therefore, the model still has limited exposure to minority-class samples, making it more difficult to learn representative semantic features and sequential patterns from the rare class.
These results demonstrate the effectiveness of minority class oversampling in improving LogLLM under imbalanced anomaly detection scenarios.

\textbf{Impact of hyperparameter $\beta$}.
As described in Section \ref{sec:mco}, the hyperparameter $\beta$ controls the proportion of the minority class by oversampling to alleviate the data imbalance problem. We investigate the impact of $\beta$ by varying its value.
Fig. \ref{fig:beta} illustrates the performance of LogLLM on the four datasets under different magnitudes of $\beta$. When  $\beta = 0$, the samples are not oversampled; instead, the original datasets are utilized directly for training.  
As illustrated in Fig. \ref{fig:RecallG}, for the HDFS, BGL, and Thunderbird datasets, the recall always increases, while for the Liberty dataset, recall decreases as $\beta$ increases. This can be attributed to the fact that for the HDFS, BGL, and Thunderbird datasets, when $\beta$ increases, anomalies are oversampled, making the model more prone to identifying samples as anomalies.
In contrast, for the Liberty dataset, when $\beta$ increases, normal samples are oversampled, making the model more prone to identifying samples as normal.
As illustrated in Fig. \ref{fig:FG}, the trend of the F$_1$-score is basically the same across all datasets. The F$_1$-score increases and then decreases as $\beta$ increases.
However, the LogLLM seems not to be sensitive to $\beta$; when $\beta$ is between 10\% and 80\%, the variation in the F$_1$-score is no more than 0.07. 
Thanks to the substantial semantic knowledge embedded in LLMs, a trained model can effectively learn anomalous patterns and detect anomalies, even when the minority class constitutes only 10\% of the dataset. 
However, LogLLM appears unable to effectively handle extremely imbalanced scenarios. For instance, in the Thunderbird dataset, anomalies constitute only 1.05\% of the samples, causing  the trained model to be biased and classify all samples as normal.  As a result, precision, recall, and F$_1$-score are all equal to 0.
Compared to the BGL and Thunderbird datasets, the precision, recall and F$_1$-score for the HDFS and Liberty datasets exhibit minimal variation with respect to $\beta$.
This consistency arises from the more distinct patterns between abnormal and normal samples in the HDFS and Liberty datasets, allowing LogLLM to easily differentiate them, regardless of the ratio of normal and abnormal samples.
As anticipated, as $\beta$ increases, the training time also increases, as shown in Fig. \ref{fig:tt}. This relationship arises because a higher  $\beta$ leads to  more oversampled data samples, as indicated by equation (\ref{eq:oversample}), thereby  enlarging the training dataset.

To summarize, minority class oversampling is essential. However, the value of the hyperparameter  $\beta$ does not significantly impact the performance of LogLLM, making careful selection unnecessary. Moreover, excessively large values of $\beta$  are undesirable, as they result in prolonged training times. 
Values between 30\% and 50\%  are deemed acceptable.

\section{Practical Deployment}
In real-world environments, LogLLM can first be trained following the methodology described in this paper and then deployed in either offline or online settings.

For offline deployment, LogLLM can serve as a diagnostic tool for postmortem analysis, where historical logs are analyzed using the same detection pipeline described in this paper. 

For online deployment, LogLLM can be integrated with existing log management platforms. Since log messages typically arrive sequentially in streaming systems, incoming messages can be accumulated until the predefined window size is reached, after which LogLLM performs anomaly detection on the constructed log sequence. 
As shown in Section \ref{sec:exp_RQ1}.\textit{Runtime comparison}, LogLLM can complete inference for one log sequence before the next-window sequence arrives, suggesting its potential applicability in online scenarios.

In both settings, batching, caching the embedding vectors of RE-preprocessed log messages, and reusing these embedding vectors can further reduce inference latency and computational cost.

\section{Threats to Validity}
In this section, we discuss the potential threats to the validity of our study.

\textbf{Generalizability to diverse industrial environments}. While LogLLM was evaluated on four widely recognized public datasets (HDFS, BGL, Liberty, and Thunderbird), these benchmarks may not fully capture the extreme noise, frequent template evolution, and complex execution flows present in modern, large-scale industrial systems. Consequently, the framework's performance might vary when applied to proprietary environments where log formats are significantly different or where anomaly patterns are more domain-specific than those captured in our experimental setup.

\textbf{Dependency on preprocessing quality and expertise}. LogLLM adopts regular expressions for preprocessing instead of automated log parsers to preserve informative semantic patterns and handle out-of-vocabulary issues. However, the effectiveness of the model is inherently tied to the quality of the regex rules used. In practice, defining precise rules for highly complex or rapidly changing log structures requires certain domain expertise, and suboptimal rules could lead to the loss of critical runtime information or the introduction of misleading semantic noise.

\textbf{Computational overhead and real-time constraints}.
The integration of LLMs like BERT and Llama provides superior semantic comprehension but results in higher memory consumption and inference latency compared to traditional lightweight models such as CNNs or LSTMs. Although we optimized the training process using QLoRA and a three-stage procedure, the inherent computational demands of LLMs may still limit the framework's suitability for scenarios requiring instantaneous, high-throughput anomaly detection on resource-constrained edge devices. Therefore, practitioners need to carefully evaluate the trade-off between detection performance, robustness, and resource requirements when applying LogLLM in different scenarios.

\textbf{Model freshness under evolving log environments}. 
Although we adopt chronological splitting to simulate realistic deployment scenarios, log environments may continue to evolve after model training, potentially causing performance degradation over time. 
To quantify the extent of log evolution, we conduct an additional analysis on the Liberty dataset. Specifically, we train the Drain log parser using logs collected from January 1, 2005 to July 1, 2005 and evaluate whether subsequent logs can be matched with the learned templates. 
The proportions of unmatched log messages are 0.014\% in July 2005, 0.006\% in August 2005, 49.300\% in September 2005, and 85.665\% in October 2005.
These results indicate significant template evolution over time, which may affect the effectiveness of models trained on historical logs.
Therefore, continuously evolving log environments may require periodic model updating or adaptation mechanisms to maintain detection performance over time.

\section{Related Work}
\label{sec:Relatedworks}
In this section, we explore related work in the field of log-based anomaly detection, with a particular focus on deep learning-based methods. 

\subsection{Traditional Deep Learning for Log-based Anomaly Detection}
Many traditional deep learning-based methods for log-based anomaly detection have been proposed.  These works can be grouped into two types based on the training paradigm: reconstruction-based methods and binary classification-based methods.

\textbf{Reconstruction-based methods} \cite{du2017deeplog,meng2019loganomaly,zhang2021logattn, catillo2022autolog, xie2023log, zhang2023anomaly,duan2021generative,he2023graph} involve designing and training a deep neural network to reconstruct input log sequences. 
Anomalies are detected based on reconstruction errors. Normal log sequences can be reconstructed with minimal errors, while anomalous log sequences cannot be effectively reconstructed, resulting in significantly higher reconstruction errors.
These methods consistently train the deep model on normal data that is free of anomalies, which means they are semi-supervised. 
Specifically, DeepLog \cite{du2017deeplog} adopts LSTM to predict the next log template ID based on past log sequences.
Similarly, LogAnomaly \cite{meng2019loganomaly}   predicts the next log template ID based on  both sequential and quantitative patterns.
Autoencoders (AEs) \cite{zhang2021logattn, catillo2022autolog, xie2023log, zhang2023anomaly} and Generative Adversarial Networks (GANs) \cite{duan2021generative,he2023graph} are widely used in reconstruction-based methods.  For example, LogAttn \cite{zhang2021logattn} adopts an AE that incorporates a Temporal Convolutional Network (TCN) to capture temporal semantic correlations and a deep neural network (DNN) to capture statistical correlations. Duan et al. \cite{duan2021generative} use a GAN, where an encoder-decoder framework based on LSTM serves as the generator. Convolutional Neural Networks (CNNs)  are used as the discriminator. The reconstruction error is calculated based on the difference between the input and the output from the generator.

\textbf{Binary classification-based methods} \cite{zhang2023layerlog,hashemi2024onelog, lu2018detecting,zhang2019robust,xie2022loggd, yilmaz2025novel, zhao2022trine,yang2021semi,studiawan2020anomaly} often employ deep neural networks that output either one or two values.  Typically, a single value represents the probability that a sample belongs to the anomalous class, and anomalies are detected by applying a threshold to convert this probability into a binary classification. When two values are output, they represent the probabilities of the sample belonging to the normal and anomalous classes, respectively.
Most methods \cite{zhang2023layerlog,hashemi2024onelog, lu2018detecting,zhang2019robust,xie2022loggd,yilmaz2025novel,studiawan2020anomaly} typically train deep models in a supervised manner. For example,
Zhang et al. \cite{zhang2023layerlog} propose LayerLog, which integrates word, log, and logseq layers to extract semantic features from log sequences. 
CNNs are utilized in \cite{hashemi2024onelog, lu2018detecting} to develop a binary classifier. LogRobust \cite{zhang2019robust} integrates a pre-trained Word2Vec model, specifically FastText \cite{joulin2016fasttext}, and combines it with TF-IDF weights to learn representation vectors of log templates. These vectors are then fed into an attention-based Bi-LSTM model for anomaly detection.
Xie et al. \cite{xie2022loggd} and Yılmaz et al. \cite{yilmaz2025novel} convert log sequences into graph structures and employ GNNs for log-based anomaly detection.
Some work \cite{zhao2022trine,yang2021semi} involves training binary classifiers in a semi-supervised manner. For example, Trine \cite{zhao2022trine} uses a Transformer encoder \cite{vaswani2017attention} to encode normal log sequences into vector representations and a generator to produce random fake vector representations. The discriminator, which is composed of a Transformer and a Multi-Layer Perceptron (MLP), is trained to distinguish whether the given vector representations are normal log sequences and it is subsequently used to detect anomalies. PLELog  \cite{yang2021semi} tackles the challenge of insufficient labeling by employing probabilistic label estimation and develops an attention-based GRU neural network for anomaly detection.

It is acknowledged that log messages are recorded in natural language and inherently carry rich semantic information. However, traditional deep learning methods often represent them as discrete IDs, thereby overlooking the semantic information. Consequently, they are less robust to unseen or evolving log templates.

\subsection{LLMs for Log-based Anomaly Detection}
\label{sec:llmad}
LLMs enable the capture of semantic and contextual patterns within system logs, facilitating more effective anomaly detection.
Existing LLM-based log anomaly detection methods can be broadly categorized into two groups: training-free methods and fine-tuning-based methods.

\textbf{Training-free methods} \cite{liu2024logprompt,qi2023loggpt,egersdoerfer2023early,pan2024raglog,zhang2025xraglog,duan2025eagerlog, zhang2025scalalog,sui2025bridging,no2024training} leverage the capabilities of LLMs without updating their parameters. 
Most training-free methods rely on prompt engineering.
For instance, Qi et al. \cite{qi2023loggpt} employ ChatGPT for zero-shot and few-shot log-based anomaly detection, utilizing prompt templates that integrate the log sequence directly. However, this approach becomes impractical when using a large window size for grouping log messages.
Egersdoerfer et al. \cite{egersdoerfer2023early} address this issue by maintaining a summary-based memory, which summarizes the previous log messages, eliminating the need to input the entire log sequence for anomaly detection.
Retrieval-Augmented Generation (RAG) is adopted in \cite{pan2024raglog,duan2025eagerlog, zhang2025scalalog, zhang2025xraglog}   for anomaly detection. For example,  RAGLog \cite{pan2024raglog} uses RAG to analyze log messages by querying its store of samples of normal log messages. They design prompt templates for LLMs to determine whether a queried log message is normal or abnormal.
Sui et al. \cite{sui2025bridging} adopt an LLM to convert each log message into a more compact sentence, bridging the syntactic gap between systems.
Instead of relying on LLM reasoning through prompts, RAPID \cite{no2024training} directly utilizes a frozen Transformer encoder-based LLM to extract semantic vectors and performs anomaly detection by comparing each query log sequence with its nearest documented counterpart.
Although training-free methods eliminate the need for model adaptation, their ability to capture dataset-specific characteristics is often limited, which may lead to suboptimal detection performance—particularly when the data distribution deviates from the knowledge encoded in LLMs.

\textbf{Fine-tuning-based methods} \cite{guo2021logbert, lee2023lanobert,lin2024fastlogad,almodovar2024logfit,le2021log,  chen2022bert, adeba2024sarlog,fu2023mlog, chu2025anomaly,hadadi2024anomaly,yang2025logllama,chai2024log} incorporate LLMs into deep neural networks and customize them to the user's own dataset. 
Some methods \cite{guo2021logbert, lee2023lanobert, lin2024fastlogad,almodovar2024logfit,yang2025logllama,chai2024log}, although adopting LLMs for anomaly detection, do not capture the semantic information within log sequences. For example, LogBERT \cite{guo2021logbert} and LAnoBERT \cite{lee2023lanobert}  utilize BERT to reconstruct the input sequence of log template IDs  (IDs of log string templates) and detect anomalies based on reconstruction errors, disregarding the semantic information.
LogLLaMA \cite{yang2025logllama} feeds the sequence of log template IDs into Llama, which is fine-tuned through a pretraining step followed by Reinforcement Learning (RL), to predict the next template ID and detect anomalies based on whether the actual template ID appears in the Top-K predicted template IDs.
Other methods \cite{le2021log, chen2022bert, adeba2024sarlog,fu2023mlog, chu2025anomaly} use Transformer encoder-based LLMs solely for extracting semantic information from log messages and then employ  smaller models \cite{le2021log, chen2022bert, adeba2024sarlog,fu2023mlog,chu2025anomaly} for classification. For instance, NeuralLog \cite{le2021log} leverages BERT to extract semantic vectors from raw log messages, which are subsequently used to detect anomalies via a Transformer-based classification model. 
Hadadi et al. \cite{hadadi2024anomaly} directly input template sequences parsed from log sequences into GPT and fine-tune the model to accurately predict sequence labels.
However, this approach faces two key challenges. First, the boundaries between templates within the sequences are unclear, making it difficult for the model to learn the sequential dependencies. Second, each template may be tokenized into multiple tokens by the LLM's tokenizer, and a single sequence can contain numerous log templates. As a result, an excessive number of tokens may be generated, often exceeding the token (memory) limits of LLMs \cite{burtsev2024working}, thereby restricting the length of sequences that can be processed.  These two challenges are further demonstrated in Section \ref{sec:ch}.

In contrast, LogLLM employs a Transformer encoder-based LLM, along with a projector, to encode each log message into a distinct embedding vector before passing it to a Transformer decoder-based LLM, thereby preserving clear boundaries between messages.
This design not only enhances the learning of sequential dependencies but also, given the same computational memory and parameter size of the Transformer decoder-based LLM, enables LogLLM to process longer log sequences than would be feasible by directly inputting the entire raw log sequence into the Transformer decoder-based LLM.

\section{Conclusion and Future Work}
\label{sec:conclusions}
In this paper, we propose LogLLM, a framework that enables collaboration between heterogeneous LLMs for log-based anomaly detection. 
LogLLM exploits the complementary capabilities of different LLM architectures, where the Transformer encoder-based LLM extracts semantic representations from individual log messages and the Transformer decoder-based LLM models sequential dependencies to generate anomaly detection decisions. 
To facilitate effective collaboration between heterogeneous LLMs, we introduce a learnable projector that aligns their representation spaces. 
Furthermore, we design a progressive three-stage training strategy to optimize the collaboration between heterogeneous LLMs and adapt the integrated framework to log anomaly detection.
Extensive experiments on four public real-world datasets demonstrate that LogLLM achieves superior performance compared with state-of-the-art methods. 
Further analyses validate the effectiveness of the key components of the model architecture and the progressive three-stage training strategy.

In future work, we plan to investigate privacy-preserving collaborative training of LogLLM across multiple software systems to enhance its robustness and generalization in heterogeneous and evolving operational environments.



\normalem 
\bibliographystyle{IEEEtran}
\bibliography{mybibliography}

\end{document}